\documentclass[aps,amsmath,amssymb,twocolumn,longbibliography,10pt,prx,superscriptaddress]{revtex4-2}
\usepackage{graphicx}
\usepackage[colorlinks=true,linkcolor=blue,citecolor=blue,urlcolor=blue]{hyperref}
\usepackage{color}
\usepackage{multirow}
\usepackage{tabularx}
\newcommand\Tstrut{\rule{0pt}{2.6ex}}
\newcommand\Bstrut{\rule[-0.9ex]{0pt}{0pt}}
\begin{document}

\title{Quantum simulation of the central spin model with a Rydberg atom and polar molecules in optical tweezers}
\author{Jacek Dobrzyniecki}
\email{Jacek.Dobrzyniecki@fuw.edu.pl}
\author{Micha{\l} Tomza}
\email{Michal.Tomza@fuw.edu.pl}
\affiliation{Faculty of Physics, University of Warsaw, Pasteura 5, 02-093 Warsaw, Poland}

\date{\today}

\begin{abstract}
Central spin models, where a single spinful particle interacts with a spin environment, find wide application in quantum information technology and can be used to describe, \emph{e.g.}, the decoherence of a qubit over time. We propose a method of realizing an ultracold quantum simulator of a central spin model with $XX$ (spin-exchanging) interactions. The proposed system consists of a single Rydberg atom (``central spin'') and surrounding polar molecules (``bath spins''), coupled to each other via dipole-dipole interactions. By mapping internal particle states to spin states, spin-exchanging interactions can be simulated. As an example system geometry, we consider a ring-shaped arrangement of bath spins, and show how it allows us to exact precise control over the interaction strengths. We numerically analyze two example dynamical scenarios which can be simulated in this setup: a decay of central spin polarization, which can represent qubit decoherence in a disordered environment, and a transfer of an input spin state to a specific output spin, which can represent the transmission of a single bit across a quantum network. We demonstrate that this setup allows us to realize a central spin model with highly tunable parameters and geometry, for applications in quantum science and technology.
\end{abstract}

\maketitle

\section{Introduction}

\label{sec:Intro}

The model of a single central spin, interacting with an environment of bath spins, is of significant scientific interest. In the field of quantum information, such central spin models can be used to describe a spin qubit immersed in a spin-bath environment, \emph{e.g.}, an electron in a semiconductor quantum dot~\cite{2002-Khaetskii-PRL,2002-Schliemann-PRB,2002-Merkulov-PRB,2003-Khaetskii-PRB,2003-Schliemann-JPhysCondMat,2004-Coish-PRB,2006-Deng-PRB,2007-Hanson-RevModPhys,2008-Yang-PRB,2009-Yang-PRB,2009-Cywinski-PRB,2009-Cywinski-PRL,2011-Cywinski-Acta,2013-Urbaszek-RevModPhys,2014-VanDenBerg-PRB,2016-Yang-RepProgPhys} or a nitrogen-vacancy center in diamond~\cite{2006-Childress-Science,2008-Hanson-Science,2011-Laraoui-PRB,2011-Zhao-PRL,2012-Zhao-PRB,2013-London-PRL,2014-Hall-PRB,2018-Schwartz-SciAdv}. In that case the model is commonly applied to analyze the decoherence of the qubit caused by its coupling to a disordered environment. Alternately, central spin models can be used to represent networks of qubits connected in a spin star topology~\cite{2004-Hutton-Arxiv,2010-Arshed-PRA,2010-Chen-PRA,2010-Anza-JPB,2011-Militello-PRA,2013-Wang-EPJD,2019-Radhakrishnan-PRA,2021-Haddadi-SciRep}, which have been analyzed for applications such as secure remote quantum computation~\cite{2018-Tran-Arxiv}, qubit state transfer~\cite{2008-HongLiang-JPB,2011-Yung-JPB}, approximate quantum state cloning~\cite{2005-DeChiara-PRA}, implementing quantum algorithms~\cite{2021-Anikeeva-PRXQ}, or generating entangled states for use in quantum communication~\cite{2004-Hutton-PRA}.

In recent decades, quantum simulators realized by ultracold particle setups have opened new possibilities of studying fundamental models of this kind. Quantum simulators allow to directly implement desired models with a high degree of control over the model parameters. This is possible thanks to developments in experimental techniques, such as optical tweezer setups which allow us to arrange individual particles into nearly-arbitrary geometries~\cite{2019-Anderegg-Science,2022-Jenkins-PRX,2022-Bluvstein-Nature,2022-Zhang-QSciTech}.

Recently, there has been a number of proposals for simulating spin models with systems of ultracold molecules~\cite{2006-Micheli-NatPhys,2007-Brennen-NJP,2011-Kuns-PRA,2011-Gorshkov-PRL,2011-Gorshkov-PRA,2012-Yao-PRL,2013-Yan-Nature,2013-Manmana-PRB,2013-Gorshkov-MolPhys,2015-Wall-NJP,2022-Li-Nature}. These proposals involve encoding spin states in molecular rotational levels, and exploiting interparticle dipolar interactions to realize effective spin-spin interactions. For example, spin-exchange processes can be simulated by exchange of excitations between molecules. Additionally, an interesting direction is combining such molecular systems with Rydberg atoms, to exploit Rydberg atom properties such as strong electric polarizability or Rydberg-blockade effects. Several example proposals for such hybrid systems exist already, in which Rydberg atoms are used to mediate interactions between molecules~\cite{2011-Kuznetsova-PCCP,2018-Kuznetsova-PRA,2022-Wang-PRXQuantum,2022-Zhang-PRXQ} or to read out molecular rotational states~\cite{2011-Kuznetsova-PCCP,2016-Kuznetsova-PRA,2022-Wang-PRXQuantum,2022-Zhang-PRXQ}. In a recent experiment, a system of a Rydberg atom and a single polar molecule in optical tweezers was created, demonstrating the first step towards experimental implementation of such systems~\cite{2023-Guttridge-PRL}.

However, in each of the mentioned proposals, the Rydberg atoms are treated as an auxiliary system, separate from the actual system of qubits encoded in polar molecules. A less explored direction are systems where Rydberg atoms and molecules are treated on equal footing, which allows us to realize two-species spin systems. The central spin model can be seen as such a two-species system, since its particle are divided into the central spin and the bath spins.

In this paper, building upon the earlier proposals for molecule-based spin-model implementations, we propose a quantum simulator realizing a central spin model. The system is composed of a single Rydberg atom, playing the role of a central spin, and a number of polar molecules, playing the role of bath spins. The setup realizes an effective spin Hamiltonian, where the particles' internal states represent levels of 1/2-spin particles, while dipole-dipole interactions between the atom and the molecules map to effective spin-spin interactions. An external magnetic or electric field can be used to tune the atomic transition between pseudospin states into resonance with a molecular transition, enabling resonant ``spin exchange'' between the particles.

Rydberg atoms are particularly advantageous for this purpose because their transition frequencies can be easily tuned even with small fields. The use of Rydberg atoms also provides an additional benefit, due to their significantly larger dipole moments compared with molecular dipole moments. As a result of this, the effective bath-bath spin interactions are weak in comparison to the effective central-bath interactions, and can therefore be neglected. This is significant, since most investigations of central spin models focus on models with vanishing bath-bath interactions.

In weak electric fields, molecules have only negligible space-fixed dipole moments, so diagonal interaction terms (corresponding to effective $z$-$z$ central-bath spin interactions) are negligible. The above properties result in the realization of an effective $XX$ central spin model, where the only interactions considered are spin-exchanging couplings between the central spin and each of the bath spins. The experimental realization of the proposed setup should be feasible in present ultracold physics laboratories.

To more fully explore the different parameter regimes of the simulated Hamiltonian, we suggest combining this setup with a particular tweezer geometry, where the molecules are arranged on a ring around the central spin. The ring can be effectively tilted by changing the direction of the applied external field, which modifies the inhomogeneity of the bath-central spin couplings. This allows us to realize both homogeneous models, where all the bath particles interact with the central spin with equal strength, and inhomogeneous models. Both these cases show distinct spin dynamics and are subjects of interest in the literature on central spin models.

To demonstrate applications of the proposed ring setup, we numerically analyze the system's time evolution in two example scenarios. In the first scenario, we simulate the decoherence of a qubit in a disordered environment, by analyzing the evolution of an initially polarized central spin. The central spin polarization decays over time, and the timescale of this decay, depending on the coupling strengths, can be regulated by changing the ring angle. In the second scenario, we simulate a transmission of a single bit across a quantum network. We set an input bath spin to an up-spin or down-spin state initially and find that over time the state of this input spin is transferred to a specific output bath spin which has the same interaction strength. The transfer timescale can be controlled by setting the ring angle.

While the presented model is relatively simple, it can serve as a demonstration of the capabilities of quantum simulators which combine atoms with polar molecules. In light of the increasing interest in such systems, and future experimental realization, it is valuable to explore the possibilities offered by this approach. Furthermore, this approach can be extended to larger systems, including systems that are not easily tractable numerically and can especially benefit from experimental quantum simulation.

This paper is organized as follows. In Sec.~\ref{sec:Model}, we describe the proposed setup and derive the effective spin Hamiltonian. In Sec.~\ref{sec:Geometry}, we consider a ring-shaped particle arrangement, and we show how this setup allows us to smoothly tune the inhomogeneity of interactions. In Sec.~\ref{sec:Dynamics}, we demonstrate the effective spin dynamics, showing how the tuneable inhomogeneity of interactions affects the resulting dynamics. Sec.~\ref{sec:Experimental} discusses more deeply various experimental aspects of this setup. Sec.~\ref{sec:Conclusion} is the conclusion.

\section{The model}

\label{sec:Model}

\subsection{The atom-molecule system}

\label{sec:Model-System}

\begin{figure}[t]
 \centering
 \includegraphics[width=0.45\textwidth]{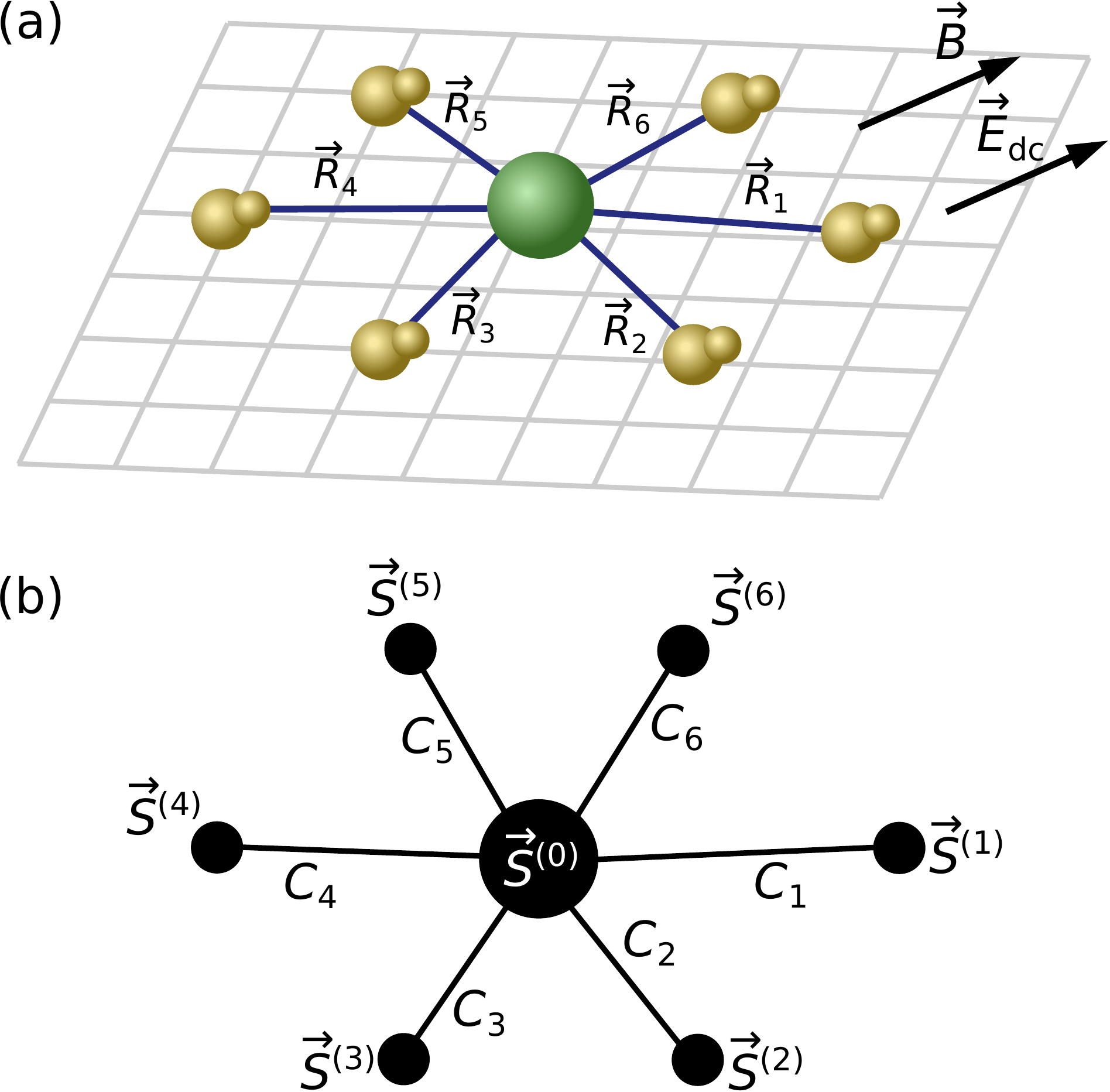}
 \caption{(a) An example $N_\mathrm{bath}=6$ atom-molecule system trapped, \emph{e.g.}, in optical tweezers. Each molecule $k=1,\ldots,N_\mathrm{bath}$ is placed at a position $\vec{R}_k$ from the atom at the center. An external electric (magnetic) field $\vec{E}_\mathrm{dc}$ ($\vec{B}$) is used to tune the particle transition frequencies as needed, and determines the space-fixed quantization axis. The depiction of molecules is schematic; in reality, typically we assume a weak electric field, where molecules do not display significant orientation along any direction. (b) A schematic depiction of the central spin model realized by the above setup. The atom acts as the central spin $\vec{S}^{(0)}$ and each molecule $k$ acts as a bath spin $\vec{S}^{(k)}$, which interacts with the central spin with strength $C_k$.}
 \label{fig:central-spin-diagram}
\end{figure}

The considered system [Fig.~\ref{fig:central-spin-diagram}(a)] consists of a single Rydberg atom (``central spin'') and $N_\mathrm{bath}$ surrounding polar molecules (``bath spins''). The particles are treated as fixed in their positions, with each molecule $k = 1,\ldots,N_\mathrm{bath}$ placed at a position $\vec{R}_k$ relative to the atom. We assume the trapping is tight enough that the extent of the particle wave functions is much smaller than the interparticle distances. Therefore we treat the particles as pointlike objects, ignoring their motional or trap states.

Such a system can be realized experimentally with optical tweezer setups, which allow to arrange particles into desired geometries. In this paper, the atom-molecule distances are taken to be $|R_k| = 1.5\,\mathrm{\mu m}$ for all $k$. This is consistent with limitations on minimum separation of tweezer traps, which is determined roughly by the wavelength of the trapping light.

The system is also subjected to an external spatially uniform dc electric field $\vec{E}_\mathrm{dc}$, and/or an external magnetic field $\vec{B}$ (with $\vec{E}_\mathrm{dc} \parallel \vec{B}$ if both are present). Their main purpose is tuning a specific atomic transition into resonance with a molecular transition, which enables the simulation of spin-exchange interactions between the atom and the molecules. Additionally, the fields determine the direction of a space-fixed quantization axis.

The Hamiltonian of the system can be written as
\begin{equation}
\label{eq:hamiltonian_initial}
 \hat{H} = \hat{h}_\mathrm{Ryd} + \sum_{k=1}^{N_\mathrm{bath}} \hat{h}_\mathrm{mol}^{(k)} + \sum_{k=1}^{N_\mathrm{bath}} \hat{V}^{(k)}_\mathrm{atom-mol}.
\end{equation}
The terms $\hat{h}_\mathrm{Ryd}$ and $\hat{h}_\mathrm{mol}^{(k)}$ describe the single-particle internal states of the Rydberg atom and molecules $k = 1,\ldots,N_\mathrm{bath}$, respectively, also taking into account the particle interactions with external fields. The last term describes electric dipole-dipole interactions between the atom and the molecules. Interactions between the molecules are neglected, since they are weak compared with the atom-molecule interactions (all energies throughout this paper are expressed in $\hbar \times 2\pi \times \mathrm{Hz}$; for simplicity, we take $\hbar = 1$).

We now examine each term of the Hamiltonian in Eq.~\eqref{eq:hamiltonian_initial} in detail. First we examine the electric dipole-dipole interaction $\sum\limits_{k} \hat{V}^{(k)}_\mathrm{atom-mol}$. The term describing an interaction between the atom and the molecule $k=1,\ldots,N_\mathrm{bath}$ is

\begin{equation}
\label{eq:interaction-k}
 \hat{V}^{(k)}_\mathrm{atom-mol} = \frac{1}{4 \pi \epsilon_0} \frac{\hat{\vec{D}} \cdot \hat{\vec{d}}^{(k)} - 3\left(\hat{\vec{D}} \cdot \vec{e}_{k}\right)\left(\hat{\vec{d}}^{(k)} \cdot \vec{e}_{k}\right)}{|R_{k}|^3}.
\end{equation}
Here $\vec{e}_{k} = \vec{R}_{k}/|\vec{R}_{k}|$, while $\hat{\vec{D}}$ ($\hat{\vec{d}}^{(k)}$) is the electric dipole moment operator acting on the atom (on the molecule $k$). The operator components $\hat{D}_q,\hat{d}^{(k)}_q$ can be written in terms of the spherical coordinate system, where basis vectors are labeled by $q = 0,\pm1$: $\vec{e}_0 \equiv \vec{e}_Z, \vec{e}_{\pm 1} \equiv \mp (\vec{e}_X \pm i \vec{e}_Y) / \sqrt{2}$ (we refer to spatial coordinates $X, Y, Z$ with capital letters, to avoid confusion with the effective spin components $x,y,z$ in the simulated spin model). The quantization axis $\vec{e}_0$ is defined to be parallel to the externally applied fields, so that $\vec{E}_\mathrm{dc} = E_\mathrm{dc} \vec{e}_0, \vec{B} = B \vec{e}_0$.

The interaction in Eq.~\eqref{eq:interaction-k} can be rewritten as a scalar product of two second-rank tensors, which correspond, respectively, to spherical harmonics and to the tensor product of dipole moment operators~\cite{2003-Brown-Book}. Omitting the complicated calculations, the result can be written out as follows:

 \begin{equation}
 \label{eq:dipole-dipole-short-form}
 \hat{V}^{(k)}_\mathrm{atom-mol} = \sum_{q',q=-1}^{+1} v^{(k)}_{q';q} \hat{D}_{q'} \hat{d}^{(k)}_q,
 \end{equation}
 where
 \begin{align}
  v^{(k)}_{0;0} = 2 v^{(k)}_{+1;-1} = 2 v^{(k)}_{-1;+1} &= \frac{1-3 \cos^2 \theta_k}{4 \pi \epsilon_0 |R_k|^3}, \\
  v^{(k)}_{0;\pm 1} = v^{(k)}_{\pm1;0} &= \frac{ \pm \frac{3}{\sqrt{2}}\sin\theta_k\cos\theta_k e^{\mp i\phi_k}} {4 \pi \epsilon_0 |R_k|^3}, \\
  v^{(k)}_{\pm 1; \pm 1} &= \frac{ -\frac{3}{2} \sin^2\theta_k e^{\mp i2\phi_k} } {4 \pi \epsilon_0 |R_k|^3}.
 \end{align}
Here $\theta_k$ is the polar angle of $\vec{R}_k$ ($\cos \theta_k = \vec{R}_k\cdot \vec{e}_0/|R_k|$) and $\phi_k$ is the azimuthal angle of rotation about the $\vec{e}_0$ axis. Note that $\hat{V}^{(k)}_\mathrm{atom-mol}$ is unaffected by choice of the X axis, as choosing a different X axis amounts to transforming the angles as $\phi_k \rightarrow \phi_k + \delta \phi$ and, simultaneously, transforming the dipole operators as $\hat{D}_{\pm 1},\hat{d}_{\pm 1} \rightarrow \hat{D}_{\pm 1} \exp(\pm i \delta \phi),\hat{d}_{\pm 1} \exp(\pm i \delta \phi)$.

We next examine the Hamiltonian term for the atom, $\hat{h}_\mathrm{Ryd}$. We assume an alkali-metal atom, for which the term can be written as
\begin{align}
\label{eq:hRyd_1stver}
 \hat{h}_\mathrm{Ryd} &= \sum\limits_{n,l,j,m_j} \mathcal{E}_{n,l,j,m_j} |n,l,j,m_j\rangle\langle n,l,j,m_j| \\
 &+ \hat{h}_\mathrm{E;atom} + \hat{h}_\mathrm{B;atom}. \nonumber
\end{align}
The states ${ |n,l,j,m_j\rangle }$ are the fine-structure basis states of the internal atomic Hamiltonian in zero external field, which have energies $\mathcal{E}_{n,l,j,m_j}$ degenerate in $m_j$. They are labeled with the usual fine-structure quantum numbers, including the total angular momentum $j$ and its projection $m_j$ on the quantization axis $\vec{e}_0$. The terms $\hat{h}_\mathrm{E;atom}$, $\hat{h}_\mathrm{B;atom}$ describe the atom's interaction with the external electric and magnetic field, respectively. Details on these interaction terms can be found, \emph{e.g.}, in Ref.~\cite{1994-Gallagher-Book}, and will not be discussed at length here.

In the limit of no external fields ($\hat{h}_\mathrm{E;atom}$, $\hat{h}_\mathrm{B;atom} = 0$), the eigenstates of Eq.~\eqref{eq:hRyd_1stver} are the levels ${ |n,l,j,m_j\rangle }$. As the external fields are increased from zero, the eigenstates of Eq.~\eqref{eq:hRyd_1stver} instead become superpositions of various ${ |n,l,j,m_j\rangle }$ with different $n$, $l$, and $j$. We can write these field-dressed eigenstates as ${ |\overline{n,l,j,m_j}\rangle }$, labeling them with the quantum numbers of their adiabatic counterparts in the zero-field limit. The corresponding dressed eigenenergies, which are no longer degenerate in $m_j$, can be written as $\overline{\mathcal{E}}_{n,l,j,m_j}$. The $\hat{h}_\mathrm{Ryd}$ term can be accordingly written in a diagonal form in this field-dressed basis:
\begin{align}
 \hat{h}_\mathrm{Ryd} = \sum\limits_{n,l,j,m_j} \overline{\mathcal{E}}_{n,l,j,m_j} |\overline{n,l,j,m_j}\rangle\langle \overline{n,l,j,m_j}|. \nonumber
\end{align}
The dressed eigenenergies $\overline{\mathcal{E}}_{n,l,j,m_j}$ at given $E_\mathrm{dc}$ or $B$, and the composition of ${ |\overline{n,l,j,m_j}\rangle }$ in terms of the zero-field levels, can be obtained numerically. For this purpose, in this paper we have used the open-source \textsc{ARC}~\cite{2017-Sibalic-CompPhys} and \textsc{pairinteraction}~\cite{2017-Weber-JPB} libraries for Python. Note that, in fields parallel to $\vec{e}_0$, the dressed states ${ |\overline{n,l,j,m_j}\rangle }$ retain $m_j$ as a good quantum number, even as $n,l,j$ cease to be good quantum numbers.

The $\hat{D}_q$ operators can couple different atom levels. The most important selection rule of this coupling is $\Delta m_j = q$. The coupling also has specific selection rules in terms of $l$ and $j$~\cite{2017-Weber-JPB}, although those rules become less important at strong external fields which mix states with different $l,j$. The corresponding off-diagonal matrix element (transition dipole moment) for a coupling between two given levels can be calculated numerically, \emph{e.g.} using the libraries mentioned above. Transitions between levels with neighboring principal quantum numbers ($n \rightarrow n\pm 0,1$) are the strongest, and the corresponding dipole matrix elements can have values up to $\sim n^2 e a_0$~\cite{2008-Gallagher-AdvAMO}.

In this paper, we focus on Rydberg states with $n \sim 50$. These values of $n$ are convenient, because the atomic transition dipole moments are large enough (around $\sim 5 \times 10^3\,\mathrm{debye}$) to result in strong interactions. Meanwhile, the Rydberg electron orbit radius $\sim a_0 n^2 \sim 0.13\,\mathrm{\mu m}$ is still significantly smaller than the interparticle distances $1.5\,\mathrm{\mu m}$. This ensures that the Rydberg electron orbit does not overlap the molecule positions, and that the atom-molecule interaction can be treated as interaction between point dipoles.

Finally, let us examine the $\hat{h}_\mathrm{mol}^{(k)}$ term, which describes the internal states of molecule $k$. In this paper, we assume polar diatomic molecules in the lowest vibrational state of a ground $\Sigma$ electronic state, which have a relatively simple eigenstate spectrum. Each term $\hat{h}_\mathrm{mol}^{(k)}$ can be written as
\begin{align}
\label{eq:molecule_internal_hamiltonian}
 \hat{h}_\mathrm{mol}^{(k)} &= \sum\limits_{N,F,M_F} \mathcal{E}_{N,F,M_F} |(N,F,M_F)^{(k)}\rangle\langle (N,F,M_F)^{(k)}| \\
 &+ \hat{h}_\mathrm{E;mol}^{(k)} + \hat{h}_\mathrm{B;mol}^{(k)} \nonumber
 \end{align}
 The states ${|(N,F,M_F)^{(k)}\rangle}$ are the hyperfine-split basis states of the internal molecular Hamiltonian in zero external field, with energies $\mathcal{E}_{N,F,M_F}$. They are labeled by the good quantum numbers in zero field: the rotational angular momentum $N = 0,1,2,\ldots$, the total angular momentum $F = 0,1,2,\ldots$, and the projection of $F$ on the quantization axis, $M_F = -F, -F+1, \ldots, +F$. The term $\hat{h}_\mathrm{E;mol}^{(k)}$ ($\hat{h}_\mathrm{B;mol}^{(k)}$) describes the interaction of the molecule with a nonzero external electric (magnetic) field.

 In the zero-field limit, the eigestates of $\hat{h}_\mathrm{mol}^{(k)}$ are the basis states ${|(N,F,M_F)^{(k)}\rangle}$. In nonzero external fields, the eigenstates of $\hat{h}_\mathrm{mol}^{(k)}$ are linear combinations of molecular eigenstates with different $F$, but $N$ and $M_F$ remain good quantum numbers. We refer to the resulting field-dressed eigenstates as ${|(\overline{N,F,M_F})^{(k)}\rangle}$, labeling them after their adiabatic counterparts in the zero-field limit. The field-dressed eigenstates, and their energies $\overline{\mathcal{E}}_{N,F,M_F}$, can be found by direct numerical diagonalization of $\hat{h}_\mathrm{mol}$ (as described in Appendix~\ref{sec:CaFSpectrum}). As a concrete example, in Appendix~\ref{sec:CaFSpectrum} we write out $\hat{h}_\mathrm{mol}$ in full, for a molecule of an example species $^{40}$Ca$^{19}$F, showing how its eigenstates change with increasing external fields. Similarly as in the case of $\hat{h}_\mathrm{Ryd}$, we can write the term Eq.~\eqref{eq:molecule_internal_hamiltonian} in diagonal form in the field-dressed basis:
 \begin{align}
 \hat{h}^{(k)}_\mathrm{mol} = \sum\limits_{N,F,M_F}                                                                                                                                                                                                                                                                                                                                                                                                                                                                                                                                                                                                                                                                                                                                                                                                                                                                                                                                                                                                                                                                                                                                                                                                                                                                                                                                                                                                                                                                                                                                                                                                                                                                                                                                                                                                                                                                                                                                                                                                                                                                                                                                                                                                                                                                                                                                                                                                                                                                                                                                                                                                                                                                                                                                                                                                          \overline{\mathcal{E}}_{N,F,M_F} |(\overline{N,F,M_F})^{(k)}\rangle\langle (\overline{N,F,M_F})^{(k)}|.
 \end{align}

The $\hat{d}^{(k)}_q$ operators can couple different molecular levels. This coupling has a number of selection rules, but for our purposes, the most significant rules are $\Delta M_F = q$ and $\Delta N = \pm 1$. The associated transition dipole moments are proportional to the molecule internuclear (body-fixed) dipole moment $d$~\cite{2003-Brown-Book}; in typical species used for ultracold experiments, $d \sim 1\,\mathrm{debye}$. On the other hand, the permanent (space-fixed) dipole moments ${\langle \overline{N,F,M_F} | \hat{d}_0 |\overline{N,F,M_F}\rangle}$ are zero in the zero-field limit, and only become non-negligible in electric fields strong enough to mix levels of different $N$. In this paper, we consider only much weaker electric fields, and so we regard the permanent molecular dipole moments as zero.

Given the $\Delta N = \pm 1$ selection rule, any effective spin exchanges mediated by dipolar interactions must occur between molecular levels with different $N$. Such pairs of levels are separated by energies on the order of the molecule's rotational constant $B_\mathrm{rot}$, which dictates the largest energy scale appearing in the effective spin Hamiltonian. In typical polar molecules of interest to the ultracold community, $B_\mathrm{rot}$ ranges from several hundred megahertz (\emph{e.g.}, $2\pi \times 353\,\mathrm{MHz}$ for RbYb~\cite{2013-Tohme-ChemPhys}) to several gigahertz (\emph{e.g.}, $2\pi \times 10.3\,\mathrm{GHz}$ for CaF~\cite{1994-Anderson-APJ}). For Rydberg atom states with $n\sim50$, transition frequencies between neighboring $n$ can also be on the order of a few GHz, making it easier to tune them into resonance with the molecular transitions.

Finally, we note a few limitations imposed by using Rydberg atoms. Since Rydberg states will play the role of effective spin states, the maximum simulation time is limited by their radiative lifetimes. A rough estimate for this lifetime in the zero-temperature limit is on the order of ${ \sim n^3 \times 10^{-9} \,\mathrm{s} }$~\cite{1994-Gallagher-Book}, which for $n \sim 50$ is ${ \sim 1 \times 10^{-4} \,\mathrm{s} }$. This limits the minimal relevant energy scales, since any Hamiltonian terms $< 2\pi \times 10^4 \,\mathrm{Hz}$ are essentially irrelevant for dynamics within the simulation time.

For example, the largest matrix elements for the molecule-molecule dipole-dipole interactions are of order $d^2/(4\pi\epsilon_0(R_k-R_{k'})^3)$. For distances $R_k - R_{k'} \sim 1\,\mathrm{\mu m}$ and typical molecule transition dipole moments $d \sim1\,\mathrm{debye}$, intermolecular interaction strengths are on the order of $\sim 2\pi \times 1.5 \times 10^2\,\mathrm{Hz}$, so neglecting them in the Hamiltonian is indeed justified. On the other hand, assuming atomic transition dipole moments $\sim 10^3\,\mathrm{debye}$ and distances $R_k = 1.5\,\mathrm{\mu m}$, the atom-molecule dipole-dipole interactions can be on the order of between $\sim 2\pi \times 10^4\,\mathrm{Hz}$ and $\sim 2\pi \times 10^5\,\mathrm{Hz}$, and thus are relevant.

Using Rydberg states also limits the maximum electric field strength $E_\mathrm{dc}$, since Rydberg atoms ionize in strong electric fields. A rough estimate for the ionization threshold field is $\sim (16 n^4)^{-1}  \hbar^2/(e m_e a^3_0)$~\cite{1994-Gallagher-Book}. For $n \sim 50$ this gives a maximum electric field $E_\mathrm{dc} \sim 50\,\mathrm{V/cm}$.

\subsection{Derivation of the effective spin Hamiltonian}

\label{sec:Model-EffectiveSpinH}

Our goal is now to rewrite the Hamiltonian of  Eq.~\eqref{eq:hamiltonian_initial} into an effective 1/2-spin Hamiltonian, in which the Rydberg atom plays the role of central spin $\vec{S}^{(0)}$, while molecules $k=1,\ldots,N_\mathrm{bath}$ play the role of bath spins $\vec{S}^{(k)}$ [as depicted in Fig.~\ref{fig:central-spin-diagram}(b)]. Specifically, we will derive an effective Hamiltonian of the form
\begin{align}
\label{eq:eff-spin-hamiltonian}
 \hat{H}_\mathrm{eff} = &c_0 \hat{S}^{(0)}_z + c_S \sum_{k=1}^{N_\mathrm{bath}} \hat{S}^{(k)}_z \nonumber \\
 &+ \sum_{k=1}^{N_\mathrm{bath}} \left( C_k \hat{S}^{(0)}_+ \hat{S}^{(k)}_- + C^*_k \hat{S}^{(0)}_- \hat{S}^{(k)}_+ \right),
\end{align}
corresponding to the $XX$ central spin model. This Hamiltonian includes an effective constant Zeeman field $c_0$ ($c_S$) acting on the central (bath) spins, as well as a spin-exchange interaction between the central spin and the bath spins, with coupling constants $C_k$. The one-body terms with $\hat{S}^{(0)}_z, \hat{S}^{(k)}_z$ in Eq.~\eqref{eq:eff-spin-hamiltonian} originate from the single-particle terms $\hat{h}_\mathrm{Ryd}, \hat{h}_\mathrm{mol}^{(k)}$ of the system Hamiltonian, while the spin-exchange terms originate from the atom-molecule interactions $\hat{V}^{(k)}_\mathrm{atom-mol}$.

The $XX$ central spin model belongs to the family of $XXZ$ central spin models, which feature anisotropic interactions of the form $\hat{S}^{(k)}_x \hat{S}^{(0)}_x + \hat{S}^{(k)}_y \hat{S}^{(0)}_y + \lambda \hat{S}^{(k)}_z \hat{S}^{(0)}_z$, where $\lambda$ varies.
	The $XX$ model ($\lambda = 0$) in particular has a number of interesting properties.
	It is fully integrable, and its entire eigenspectrum can be found via Bethe ansatz methods.
	Additionally, for inhomogeneous $C_k$, it is the only $XXZ$ model to have dark eigenstates, in which the central spin is fully polarized and completely unentangled with the bath.
	Spectral properties of the $XX$ model have been studied, \emph{e.g.}, in Refs.~\cite{2020-Villazon-PRR,2020-Villazon-SciRep,2022-Dimo-PRB}, and analytical expressions for its dynamics are available for specific initial states~\cite{2009-Jivulescu-PhysScr,2009-Jivulescu-RepMatPhys}. With regard to practical applications, the $XX$ central spin model can describe, \emph{e.g.}, an electron in a semiconductor quantum dot, interacting with an external magnetic field which cancels out its $z$-$z$ interactions with the bath~\cite{2003-Taylor-PRL-Long,2014-Ding-PRB}. It can also arise in resonantly driven spin systems in rotating frames~\cite{1962-Hartmann-PR,2008-Rovnyak-ConcMagnRes,2018-FernandezAcebal-NanoLetters,2020-BhaktavatsalaRao-NJP}.

\subsubsection{Defining the effective spin states}

To simulate the 1/2-spin model in our setup, first we choose a pair of internal atom (molecule) states to act as central (bath) spin states, and we treat each particle as a two-level system. The two central spin states are denoted as ${|\Uparrow\rangle},{|\Downarrow\rangle}$, \emph{i.e.}, up and down respectively, while the two states of bath spin $k$ are denoted ${|\uparrow^{(k)}\rangle},{|\downarrow^{(k)}\rangle}$ (\emph{i.e.}, we use the superscript $(k)$ to indicate a state from the Hilbert subspace associated with particle $k$). Further in the paper, we use the following compact notation for product states: ${ |\Uparrow;\downarrow^{(k)}\rangle } \equiv {|\Uparrow\rangle \otimes |\downarrow^{(k)}\rangle }$, \emph{etc.}, and ${ |\Uparrow;\uparrow^{(1)},\downarrow^{(2)},\ldots,\uparrow^{(N)}\rangle } \equiv { |\Uparrow\rangle \otimes |\downarrow^{(1)}\rangle \otimes |\uparrow^{(2)}\rangle \otimes \ldots \otimes |\uparrow^{(N)}\rangle }$.

We denote the energies of the chosen atomic states as $\mathcal{E}_\Uparrow,\mathcal{E}_\Downarrow$, and the difference between their quantum numbers $m_j$ as $\delta m_j \equiv {m_j}_\Uparrow - {m_j}_\Downarrow$. Similarly, we denote the energies of molecular states as $\mathcal{E}_\uparrow,\mathcal{E}_\downarrow$, and the difference of their quantum numbers $M_F$ as $\delta M_F \equiv {M_F}_\uparrow - {M_F}_\downarrow$.

We can define the associated (dimensionless) identity and spin operators as
\begin{equation}
\begin{aligned}
\hat{1}^{(0)} &= |\Downarrow\rangle\langle\Downarrow|+|\Uparrow\rangle\langle\Uparrow|, \\
\hat{S}^{(0)}_z &= \frac{1}{2}|\Uparrow\rangle\langle\Uparrow| - \frac{1}{2}|\Downarrow\rangle\langle\Downarrow|, \\
\hat{S}^{(0)}_+ &= |\Uparrow\rangle\langle\Downarrow|, \\
\hat{S}^{(0)}_- &= |\Downarrow\rangle\langle\Uparrow|, \\
\hat{1}^{(k)} &= |\downarrow^{(k)}\rangle\langle\downarrow^{(k)}|+|\uparrow^{(k)}\rangle\langle\uparrow^{(k)}|, \\
\hat{S}^{(k)}_z &= \frac{1}{2}|\uparrow^{(k)}\rangle\langle\uparrow^{(k)}| - \frac{1}{2}|\downarrow^{(k)}\rangle\langle\downarrow^{(k)}|, \\
\hat{S}^{(k)}_+ &= |\uparrow^{(k)}\rangle\langle\downarrow^{(k)}|, \\
\hat{S}^{(k)}_- &= |\downarrow^{(k)}\rangle\langle\uparrow^{(k)}|
\end{aligned}
\end{equation}
(with $1 \le k \le N_\mathrm{bath}$).

There are a few criteria for choosing the pseudospin ${ \Uparrow,\Downarrow }$ and ${ \uparrow,\downarrow }$ states. First, the $V_\mathrm{atom-mol}^{(k)}$ term must be able to realize the effective spin exchange, \emph{i.e.}, the matrix elements ${ \langle \Uparrow ; \downarrow^{(k)} | \hat{V}_\mathrm{atom-mol}^{(k)} |\Downarrow ; \uparrow^{(k)} \rangle }$ must be nonzero. Therefore the atomic transition dipole moment ${ \mu_\Updownarrow = {\langle \Uparrow | \hat{D}_{\delta m_j} | \Downarrow \rangle} }$ and the molecular transition dipole moment ${ \mu_\updownarrow = {\langle \uparrow^{(k)} | \hat{d}^{(k)}_{\delta M_F} | \downarrow^{(k)} \rangle} }$ should be nonzero. Usually, $\mu_\Updownarrow, \mu_\updownarrow$ should also be as large as possible, to maximize the interaction magnitudes $|C_k|$. This is to ensure that the interaction timescales $\approx 1/|C_k|$ are much smaller than the maximum simulation time. In the rigid-rotor molecule, the strongest transition dipole moment is that which corresponds to the ${ N=0 \leftrightarrow N=1 }$ transition. Therefore, a pair of levels with $N=0, N=1$ is the optimal choice for the bath spin states. Similarly, for the atomic transition, a good choice is a $n\rightarrow n$ or $n\rightarrow n\pm 1$ transition.

The next criterion is that the spin-exchange process ${| \Uparrow ; \downarrow^{(k)} \rangle \leftrightarrow | \Downarrow ; \uparrow^{(k)} \rangle}$ must be energetically resonant. That is, its associated energy change ${ |c_\Delta| = |(\mathcal{E}_\Uparrow - \mathcal{E}_\Downarrow) - (\mathcal{E}_\uparrow - \mathcal{E}_\downarrow)| }$ should be smaller than the magnitude of the interaction matrix element ${ \langle \Uparrow; \downarrow^{(k)} | V_\mathrm{atom-mol}^{(k)} | \Downarrow; \uparrow^{(k)} \rangle }$, which is on order of ${ \sim \mu_\Updownarrow \mu_\updownarrow (4 \pi \epsilon_0 |R_k|^3)^{-1} }$.
Simultaneously, ${ \mathcal{E}_\Uparrow - \mathcal{E}_\Downarrow }$ and ${ \mathcal{E}_\uparrow - \mathcal{E}_\downarrow }$ must be large enough that processes changing the total spin, such as ${ |\Uparrow ;\downarrow\rangle \leftrightarrow |\Uparrow; \uparrow\rangle }$, are not energetically resonant. This latter condition is easily fulfilled in our system, where molecular and atomic transitions have frequencies of ${ \sim 2\pi \times 10^9\,\mathrm{Hz} }$ magnitudes, whereas the atom-molecule dipolar interactions are at most of order ${ \sim 2\pi \times 10^5\,\mathrm{Hz} }$ at the assumed interparticle distances.

The final criterion is that, once excited into the pseudospin states, the particles must stay within the pseudospin basis throughout the entire simulation. Thus, any interaction process involving a transition from a pseudospin to a non-pseudospin state must be suppressed (either disallowed by electric dipole selection rules, or energetically off-resonant).

To fulfill the criterion of a small mismatch $c_\Delta$, the external electric or magnetic field can be used to tune the atomic and molecular transition frequencies into resonance. Using a magnetic field has the advantage of lifting the degeneracies between atomic (molecular) levels with different $m_j$ ($M_F$), making it easier to avoid accidental resonant transitions to states outside the pseudospin basis. Using an electric field lifts degeneracies only between levels with different $|m_j|$ ($|M_F|$), and so it may be insufficient by itself. It is possible to combine an electric and a magnetic field, to more precisely adjust the energy spectra, at cost of greater experimental complication.

\subsubsection{Deriving the effective spin-exchange coupling}

By projecting the dipole operators $\hat{D}_q$ ($\hat{d}^{(k)}_q$) in $\hat{V}^{(k)}_\mathrm{atom-mol}$ [Eq.~\eqref{eq:dipole-dipole-short-form}] on the states ${ |\Uparrow\rangle},{|\Downarrow\rangle  }$ (${ |\uparrow^{(k)}\rangle},{|\downarrow^{(k)}\rangle }$), and neglecting all the small or off-resonant terms in $\hat{V}^{(k)}_\mathrm{atom-mol}$, the interaction can be written as
\begin{align}
\label{eq:effective-spin-interaction}
  \hat{V}^{(k)}_\mathrm{atom-mol} &= C_k |\Uparrow\rangle\langle\Downarrow| |\downarrow^{(k)}\rangle\langle\uparrow^{(k)}| + \mathrm{H.c.} \nonumber  \\
  & = C_k \hat{S}^{(0)}_+ \hat{S}^{(k)}_- + C_k^* \hat{S}^{(0)}_- \hat{S}^{(k)}_+.
\end{align}
The effective coupling constant $C_k$ is given by
\begin{equation}
\label{eq:effective-interaction-coefficient}
C_k = \langle \Uparrow ; \downarrow^{(k)} | \hat{V}_\mathrm{atom-mol}^{(k)} | \Downarrow ; \uparrow^{(k)} \rangle.
\end{equation}
More specifically,
\begin{align}
C_k &= \sum_{q,q'} v^{(k)}_{q';q} \langle \Uparrow | \hat{D}_{q'} | \Downarrow \rangle \langle \downarrow^{(k)} | \hat{d}^{(k)}_{q} | \uparrow^{(k)} \rangle \nonumber \\
&= v^{(k)}_{\delta m_j;-\delta M_F} \mu_\Updownarrow (-1)^{\delta M_F} \mu_\updownarrow.
\end{align}

For example, suppose that $\Uparrow$ and $\Downarrow$ are chosen as two atomic states with $\delta m_j = 0$, while $\uparrow$ and $\downarrow$ are chosen as two molecular levels with $\delta M_F = 0$. Then we have
\begin{align}
C_k &= v^{(k)}_{0;0} \langle \Uparrow | \hat{D}_{0} | \Downarrow \rangle \langle \downarrow^{(k)} | \hat{d}^{(k)}_{0} | \uparrow^{(k)} \rangle  \nonumber \\
&= \frac{\mu_\Updownarrow \mu_\updownarrow}{4 \pi \epsilon_0 |R_k|^3} (1-3\cos^2 \theta_k).
\end{align}
Therefore, the choice of pseudospin states determines both the overall magnitude of $C_k$ (via the values of $\mu_\updownarrow, \mu_\Updownarrow$) and its angular dependency (via the values of $\delta M_F$ and $\delta m_j$).

Significantly, the resulting effective spin interaction [Eq.~\eqref{eq:effective-spin-interaction}] does not include diagonal terms (of the form ${ \propto \hat{S}^{(0)}_z \hat{S}^{(k)}_z }$). This is because the diagonal matrix elements of the interaction, such as ${\langle \Uparrow ; \downarrow^{(k)} | V_\mathrm{atom-mol}^{(k)} | \Uparrow ; \downarrow^{(k)} \rangle}$, are proportional to the molecular permanent dipole moments, which makes them negligible.

It is important to note that $C_k$ may be negative, and even may be complex (if $\delta M_F \ne \delta m_j$). However, for the $XX$ central spin Hamiltonian, it is possible to effectively modify the phase of any $C_k$ for the purposes of time evolution, simply by appropriately modifying the initial state. This procedure is described in Appendix~\ref{sec:Phase-Transformation}.

Finally, we note that if the molecular spectrum includes degenerate levels, it is possible to define the pseudospin states as linear combinations of such levels. This can result in $C_k$ taking forms other than Eq.~\eqref{eq:effective-interaction-coefficient}. See Appendix~\ref{sec:LinearCombination} for details.

\subsubsection{Deriving the one-body effective spin terms}

The one-particle terms $\hat{h}_\mathrm{Ryd}$ and $\hat{h}_\mathrm{mol}^{(k)}$ in the Hamiltonian of Eq.~\eqref{eq:hamiltonian_initial} are straightforward to convert to effective spin operator form. Projecting them on the states ${|\Uparrow\rangle},{|\Downarrow\rangle}$ (${|\uparrow^{(k)}\rangle},{|\downarrow^{(k)}\rangle}$), we get
\begin{align}
\label{eq:atom-1particle-part-eff-spin-form}
 \hat{h}_\mathrm{Ryd} &= \mathcal{E}_\Uparrow |\Uparrow\rangle\langle \Uparrow| + \mathcal{E}_\Downarrow |\Downarrow\rangle\langle \Downarrow| \\
 &= c_0 \hat{S}^{(0)}_z + \frac{\mathcal{E}_\Uparrow + \mathcal{E}_\Downarrow}{2} \hat{1}^{(0)}, \nonumber \\
\label{eq:mol-1particle-part-eff-spin-form}
 \hat{h}_\mathrm{mol}^{(k)} &= \mathcal{E}_\uparrow |\uparrow^{(k)}\rangle\langle \uparrow^{(k)}| + \mathcal{E}_\downarrow |\downarrow^{(k)}\rangle\langle \downarrow^{(k)}| \\
 &= c_S \hat{S}^{(k)}_z + \frac{\mathcal{E}_\uparrow + \mathcal{E}_\downarrow}{2} \hat{1}^{(k)}, \nonumber
\end{align}
where the coefficients are given by
\begin{align}
 c_0 &= \mathcal{E}_\Uparrow - \mathcal{E}_\Downarrow, \label{eq:ct-coefficient} \\
 c_S &= \mathcal{E}_\uparrow - \mathcal{E}_\downarrow. \label{eq:cs-coefficient}
\end{align}

After discarding the constant parts in Eqs.~\eqref{eq:atom-1particle-part-eff-spin-form} and ~\eqref{eq:mol-1particle-part-eff-spin-form}, we finally arrive at an effective spin Hamiltonian of the form given in Eq.~\eqref{eq:eff-spin-hamiltonian}. The obtained values of the parameters $c_0, c_S, C_k$ are given by Eqs.~\eqref{eq:ct-coefficient},~\eqref{eq:cs-coefficient} and ~\eqref{eq:effective-interaction-coefficient}, respectively.

The Hamiltonian in Eq.~\eqref{eq:eff-spin-hamiltonian} conserves the total spin $z$ component, defined as
\begin{equation}
\label{eq:total-spin-z}
 \hat{\sigma}_z = \hat{S}^{(0)}_z + \sum_{k=1}^{N_\mathrm{bath}} \hat{S}^{(k)}_z,
\end{equation}
and therefore it can be also written as
\begin{equation}
\label{eq:hamiltonian_with_total_spin}
  \hat{H}_\mathrm{eff} = c_\Delta \hat{S}^{(0)}_z + c_S \hat{\sigma}_z + \sum_{k=1}^{N_\mathrm{bath}} \left( C_k \hat{S}^{(0)}_+ \hat{S}^{(k)}_- +  C^*_k \hat{S}^{(0)}_- \hat{S}^{(k)}_+ \right),
\end{equation}
where the energy mismatch $c_\Delta = c_0 - c_S$. In fact, for certain initial states, ${ \langle c_S \hat{\sigma}_z \rangle }$ is a constant. This happens if the initial state is a linear combination of product states which all have the same value of ${\langle\hat{\sigma}_z\rangle}$. In that case, the system evolution acts as if governed by the Hamiltonian
\begin{equation}
\label{eq:effective-spin-hamiltonian-constant-spin}
\hat{H}'_\mathrm{eff} = c_\Delta \hat{S}^{(0)}_z + \sum_{k=1}^{N_\mathrm{bath}} \left( C_k \hat{S}^{(0)}_+ \hat{S}^{(k)}_- +  C^*_k \hat{S}^{(0)}_- \hat{S}^{(k)}_+ \right),
\end{equation}
with an effective Zeeman field acting only on the central spin.

\subsection{Numerical dynamics simulation}

To obtain the dynamics, we simulate the time evolution of the system numerically by exact diagonalization of the Hamiltonian. First, we define a many-body basis, made up of all the possible $2^{1+N_\mathrm{bath}}$ product states ${ |S^{(0)} ; S^{(1)}, \ldots , S^{(N_\mathrm{bath})}\rangle }$ where ${ S^{(0)} \in \{ \Uparrow,\Downarrow \} }$ and ${ S^{(k)} \in \{ \uparrow^{(k)}, \downarrow^{(k)} \} }$. The Hamiltonian in Eq.~\eqref{eq:eff-spin-hamiltonian} is diagonalized in that basis (using the standard \textsc{LAPACK} diagonalization routines) to find its eigenstates ${ |K\rangle }$ and eigenergies $E_K$. Then, for an arbitrary initial state ${ |\Psi_0\rangle }$, the system state at any time $t$ is given by
 \begin{equation}
  |\Psi(t)\rangle = e^{-i \hat{H}_\mathrm{eff} t} |\Psi_0\rangle = \sum_{\{|K\rangle\}} \left( e^{-i E_K t } \langle K | \Psi_0\rangle\right) |K\rangle.
 \end{equation}

To reduce the complexity of the calculations, we exploit the fact that the total spin $z$ projection $\sigma_z$ [Eq.~\eqref{eq:total-spin-z}] is conserved. Therefore, to find the evolution of a given initial state, usually we can consider only a small part of the entire basis. For example, for an initial state ${ |\Uparrow; \downarrow_1 \ldots \downarrow_{N_\mathrm{bath}}\rangle }$, only states with one spin up (which are only $N_\mathrm{bath} + 1$ in number) need to be included in the basis. This allows us to obtain the accurate time evolution efficiently, without needing to use more complicated numerical techniques.

\subsection{Possible parameter ranges of the effective Hamiltonian}
\label{sec:Limitations}

It is useful to estimate the experimentally obtainable ranges of parameters $c_S$, $c_0$, and $C_k$. The coupling coefficients $C_k$ are dependent on $\vec{R}_k$, and can be quite freely adjusted by placing the molecules at different positions. With realistic values of transition dipole moments, $\mu_\updownarrow \sim 1\,\mathrm{debye}$ and $\mu_\Updownarrow \sim 5 \times 10^3\,\mathrm{debye}$, for atom-molecule distances $R_k = 1.5\,\mathrm{\mu m}$ the interaction strengths $\mu_\updownarrow \mu_\Updownarrow / (4 \pi \epsilon_0 |R_k|^3)$ are typically on the order of ${ 2\pi \times 10^5\,\mathrm{Hz} }$. The interaction can potentially be increased further by using molecule species with larger dipole moments, atomic transitions with larger $\mu_\Updownarrow$, or by moving the molecules closer to the atom, although the distance $|R_k|$ cannot be made arbitrarily small due to experimental limits on positioning optical tweezers and the finite size of Rydberg atoms.

In contrast, $c_0,c_S$ are mostly outside of the experimenter's direct control. They depend primarily on the frequency of the chosen molecular transition, which is determined by the rotational constant of the molecular species. The rotational constant, in principle, can be modified by exciting the molecule to higher vibrational levels, but this change is usually small, and molecular dipole moments decrease with vibrational excitations.

The values of $c_S$ and $c_0$ are typically much larger than the interaction strengths. This can cause difficulties in preserving phase relations between components of the system wave function ${ |\Psi(t)\rangle }$ that have different total spin $z$ projection ${ \langle \sigma_z \rangle }$. The phase differences between those components will oscillate with frequencies on the order $\sim c_S$. Unless the system can be controlled on timescales $1/c_S$ (typically of nanosecond order or less, much smaller than the microsecond timescales required if $C_k$ is the only energy scale), properly preserving phase information is difficult. Although this is not a fundamentally unsurmountable difficulty, it would complicate experimental implementation. This difficulty is irrelevant, however, if the initial state consists only of product states ${ |S^{(0)} ; S^{(1)}, \ldots , S^{(N_\mathrm{bath})}\rangle }$ which all have the same value of ${\langle\hat{\sigma}_z\rangle}$. Then the dynamics, being governed by the effective Hamiltonian $H'_\mathrm{eff}$ [Eq.~\eqref{eq:effective-spin-hamiltonian-constant-spin}], are only affected by the difference $c_\Delta$, which can be set to small values with external fields.

\subsection{Choice of atomic and molecular species}

\label{sec:Model-Species}

\subsubsection{Possible species of atom and molecules}

There exists a wide variety of atomic and molecular species that can be used to realize the proposed setup. For the atomic species, a highly suitable choice are alkali-metal atoms, which are relatively easy to cool to ultracold temperatures due to their favorable electronic structure. Trapping single ultracold alkali-metal atoms in optical tweezers has been extensively demonstrated~\cite{2006-Miroshnychenko-Nature,2006-Miroshnychenko-PRL,2008-Tuchendler-PRA,2012-Muldoon-NJP}.

For the molecular species, a suitable choice are molecules which can be reliably cooled to the ground state and trapped in optical tweezers. Preferrably, they should have a strong dipole moment, resulting in strong interactions. One class of such species are diatomic alkali-metal molecules (with a $^1\Sigma$ ground state), which typically have relatively strong dipole moments $d$ of several debye. Examples are LiCs with $d = 5.5\,\mathrm{debye}$ or NaCs with $d = 4.6\,\mathrm{debye}$~\cite{2005-Aymar-JChemPhys}. Alkali-metal dimers have been already successfully prepared in electronic and rovibrational ground states, as demonstrated for KRb~\cite{2010-Aikawa-PRL} and LiCs~\cite{2008-Deiglmayr-PRL}. NaCs molecules in desired internal states have been created in optical tweezers~\cite{2018-Liu-Science,2019-Liu-PRX,2020-Zhang-PRL,2021-Cairncross-PRL}, and very recently RbCs molecules were obtained as well~\cite{2023-Ruttley-PRL}.

An even more promising category are molecules in $^{2}\Sigma$ or $^{3}\Sigma$ states, which contain unpaired electron spins.
Such molecules are highly tunable via magnetic fields thanks to their high magnetic moment. Even moderate fields of a few hundred Gauss can create energy splittings large in comparison to the interaction strengths, making it easier to prevent resonant transitions to unwanted molecular states.
A prime example of such species are CaF molecules with a $^2\Sigma$ ground state, which have been cooled and trapped in optical tweezers in several experiments~\cite{2019-Anderegg-Science,2020-Cheuk-PRL,2021-Anderegg-Science,2021-Burchesky-PRL,2022-Holland-PRL,2022-Bao-Arxiv}. Other promising examples include SrF (which has been successfully cooled to ultracold temperatures~\cite{2014-Barry-Nature}) as well as dimers consisting of an alkali-metal atom combined with closed-shell alkaline-earth-metal atoms. Many of these species have the required strong electric dipole moments, for example CaF with $3.07\,\mathrm{debye}$~\cite{1984-Childs-JChemPhys} or SrF with $3.47\,\mathrm{debye}$~\cite{1985-Ernst-ChemPhysLett}.

\subsubsection{Example choice of atom and molecule pair}
\label{sec:Model-ExampleChoice}

As a specific example of an usable atom-molecule combination, here we will consider a system composed of $^{40}$Ca$^{19}$F molecules as the bath spins and a $^{39}$K atom as the central spin. Assume the electric field is set to zero, while the magnetic field is set to a particular value $B_\mathrm{res} = 818.1\,\mathrm{G}$. We designate the following field-dressed $^{40}$Ca$^{19}$F levels as bath spin states:
\begin{align}
|\downarrow^{(k)}\rangle &= |(\overline{N=0,F=0,M_F=0})^{(k)}\rangle,\\
|\uparrow^{(k)}\rangle &= |(\overline{N=1,F=1^-,M_F=0})^{(k)}\rangle
\end{align}
(there are two $N=1,F=1$ hyperfine manifolds in the $^{40}$Ca$^{19}$F spectrum; we use $F = 1^-$ to distinguish the lower-energy one). When dressed by the given magnetic field, these levels can be even more conveniently described in terms of the uncoupled molecular state basis (described in Appendix~\ref{sec:CaFSpectrum}):
\begin{align}
|\downarrow^{(k)}\rangle &\approx \left|\left(N=0,M_N=0,M_s=-\frac{1}{2},M_I=+\frac{1}{2}\right)^{(k)}\right\rangle,\\
|\uparrow^{(k)}\rangle &\approx \left|\left(N=1,M_N=0,M_s=-\frac{1}{2},M_I=+\frac{1}{2}\right)^{(k)}\right\rangle.
\end{align}
The numerically calculated molecular transition dipole moment is $\mu_\updownarrow \equiv {\langle \uparrow^{(k)} | \hat{d}^{(k)}_{0} | \downarrow^{(k)} \rangle} = d/\sqrt{3} = 1.77\,\mathrm{debye}$.

\begin{figure}[t]
 \centering
 \includegraphics[width=0.5\textwidth]{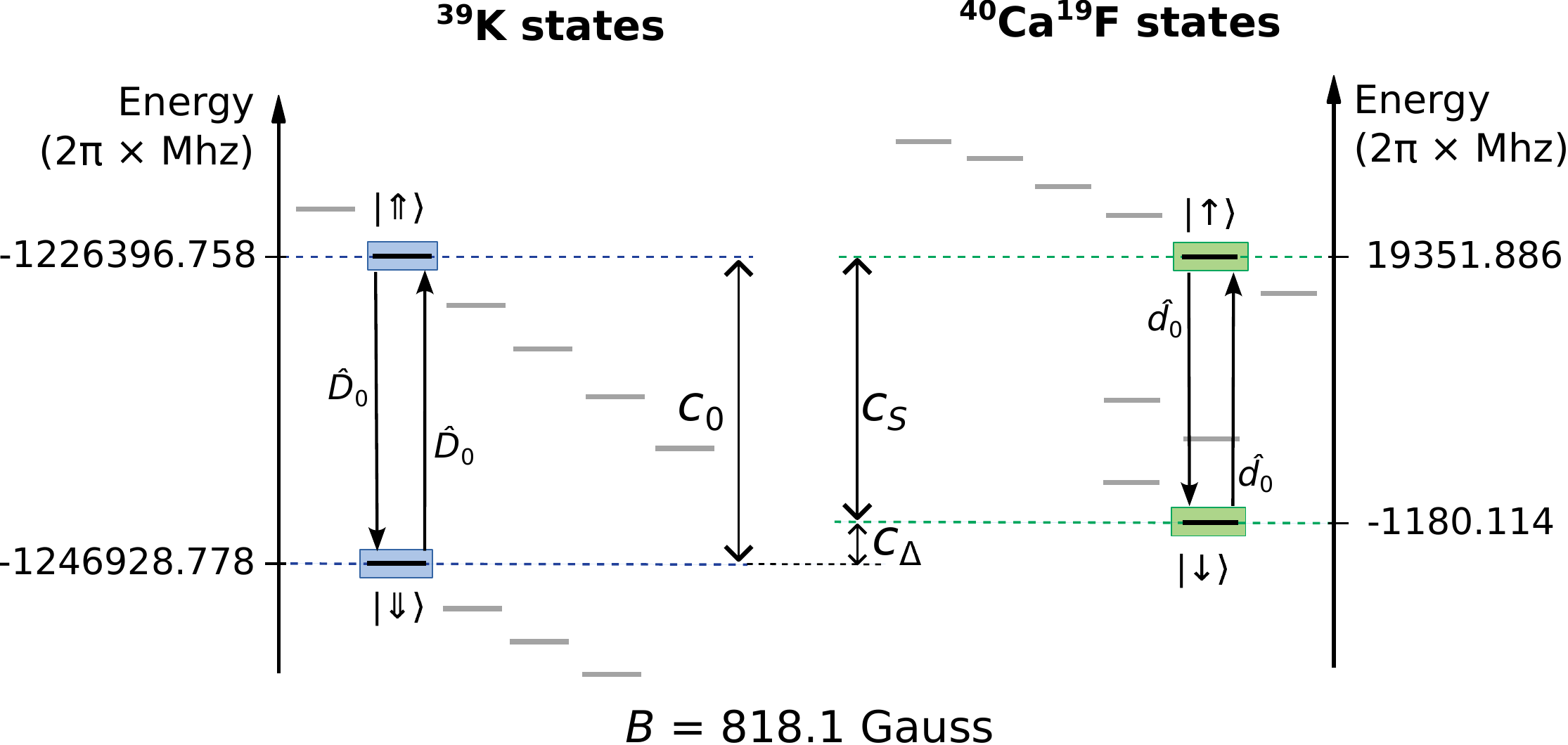}
 \caption{Left: Gray lines correspond to the ${ n=52,l=2,j=5/2 }$ and ${ n=53,l=1,j=3/2 }$ energy levels of a $^{39}$K atom in a magnetic field of $818.1\,\mathrm{G}$. Two of the energy levels (enclosed in boxes), differing in energy by $c_0$, are chosen as pseudospin states ${ |\Uparrow\rangle,|\Downarrow\rangle} $. The two levels can be coupled by the action of the electric dipole operator $\hat{D}_{0}$, as shown. The state energies are shown relative to the K atom's ionization level. Right: Gray lines correspond to the $N=0$ and $N=1$ energy levels of a $^{40}$Ca$^{19}$F molecule, in the same magnetic field. Two of the energy levels, differing in energy by $c_S$, are chosen as pseudospin states ${ |\uparrow\rangle,|\downarrow\rangle }$. The two levels can be coupled by the action of the electric dipole operator $\hat{d}_{0}$, as shown. At this magnetic field, the atomic transition energy $c_0$ and the molecular transition energy $c_S$ differ by a value $c_\Delta = 2\pi \times 19 \,\mathrm{kHz}$. (The placement of the energy levels on the vertical axis is schematic, and does not reflect actual proportions between state energies.)}
 \label{fig:AtomAndMoleculeStates}
\end{figure}

The central spin states are taken as the following atomic Rydberg levels:
\begin{align}
|\Downarrow\rangle &= \left|\overline{n=53,l=1,j=\frac{3}{2},m_j=+\frac{3}{2}}\right\rangle,\\
|\Uparrow\rangle &= \left|\overline{n=52,l=2,j=\frac{3}{2},m_j=+\frac{3}{2}}\right\rangle.
\end{align}
In the magnetic field $B_\mathrm{res}$, the $|\Downarrow\rangle$ state is unchanged from its zero-field form, while the $|\Uparrow\rangle$ state is dominated by an admixture of ${ \left|n=52,l=2,j=5/2,m_j=+3/2\right\rangle }$. The numerically calculated atomic transition dipole moment is $\mu_\Updownarrow \equiv {\langle \Uparrow | \hat{D}_{0} | \Downarrow \rangle} = 3626.87\,\mathrm{debye}$. A schematic overview of the chosen atomic and molecular pseudospin states is shown in Fig.~\ref{fig:AtomAndMoleculeStates}. The relevant parts of the atomic and molecular spectra are shown in more detail in Fig.~\ref{fig:K-CaF-Zeeman-levels} in Appendix~\ref{sec:Spectra}.

The magnetic field value $B_\mathrm{res} = 818.1\,\mathrm{G}$ is an optimal value at which the transition frequencies $c_S,c_0$ are almost equal, with a minimal residual difference $c_\Delta = 2\pi \times 19\,\mathrm{kHz}$ (as shown in Fig.~\ref{fig:AtomAndMoleculeStates}). $B_\mathrm{res}$ has been determined by numerical calculation, to within a limited accuracy of $0.1\,\mathrm{Gauss}$ (hence $c_\Delta$ has a residual nonzero value). This reflects the fact that, in real experiments, the ability to control $B$ is limited by environmental field fluctuations and other sources of error. Realistically, in experiments the magnetic field might be controllable with relative $10^{-4}$ accuracy. We have calculated that, around the value $B_\mathrm{res}$, changing $B$ by $0.1\,\mathrm{G}$ changes $c_\Delta$ by $2\pi \times 96\,\mathrm{kHz}$. This means that $10^{-4}$ accuracy suffices to keep $c_\Delta$ no larger than $\sim 2\pi \times 100\,\mathrm{kHz}$.

Assuming particle distances of $|R_k| = 1.5\,\mathrm{\mu m}$, the resulting coupling coefficients $C_k$ are given by
\begin{equation}
\label{eq:example_Ck_expression}
 C_k = \frac{\mu_\updownarrow \mu_\Updownarrow}{4 \pi \epsilon_0 |R_k|^3} (1-3 \cos\theta_k^2) = 2\pi \times 287\,\mathrm{kHz} (1-3 \cos\theta_k^2).
\end{equation}
The Rydberg state radiative lifetimes (determined numerically, in the zero-field limit) are of order $\sim 10^{-4}\,\mathrm{s}$. Compared with the inverse of these lifetimes, the interactions are reasonably strong, confirming that meaningful spin-exchange dynamics can be observed within the maximum simulation time.

\begin{table}[t]
\begin{tabular}{l@{\hskip 0.25in}l}
\hline
\hline
$B_\mathrm{res}$ & $818.1\,\mathrm{G}\dagger$ \Tstrut \Bstrut\\
\hline
$B_\mathrm{rot}$ & $2\pi \times 10267.539\,\mathrm{MHz}$~\cite{1994-Anderson-APJ} \Tstrut\\
$d$ & $3.07\,\mathrm{debye}$~\cite{1984-Childs-JChemPhys} \\
$c_S$ & $2\pi \times 20532.001\,\mathrm{MHz}$\\
$\mu_\updownarrow$ & $1.77\,\mathrm{debye}$ \Bstrut\\
\hline
$c_0$ & $2\pi \times 20532.020\,\mathrm{MHz}\dagger$ \Tstrut\\
$c_\Delta$ & $2\pi \times 19\,\mathrm{kHz}\dagger$ \\
$\mu_\Updownarrow$ & $3626.87\,\mathrm{debye}\dagger$ \Bstrut\\
\hline
$\mu_\updownarrow \mu_\Updownarrow/(4 \pi \epsilon_0 |1.5\,\mathrm{\mu m}|^3)$ & $2\pi \times 287\,\mathrm{kHz}$ \Tstrut \Bstrut\\
\hline
Lifetime of ${ |\Uparrow\rangle }$ & $3.4 \times 10^{-4}\,\mathrm{s}\ddagger$ \Tstrut\\
Lifetime of ${ |\Downarrow\rangle }$ & $5.2 \times 10^{-4}\,\mathrm{s}\ddagger$ \Bstrut\\
\hline
\hline
\end{tabular}
\caption{Relevant physical parameters of the $^{39}$K--$^{40}$Ca$^{19}$F hybrid system in a magnetic field $B = B^\mathrm{res}$, tuned to equalize the transition frequencies $c_0,c_S$. $\dagger$ marks atomic state properties obtained by a calculation with the Python \textsc{pairinteraction} package. $\ddagger$ marks atomic radiative lifetimes, obtained by a calculation with the Python \textsc{ARC} package~\cite{2017-Sibalic-CompPhys}. Radiative atomic state lifetimes are calculated for the $0\,\mathrm{K}$ temperature limit, under the assumption of no electronic or magnetic fields.}
\label{tab:CaFKparameters}
\end{table}

In Table~\ref{tab:CaFKparameters} the physical properties of the system are listed, as well as the resulting parameters that appear in the effective spin Hamiltonian. We use these parameters in the example dynamical simulations in later sections.

\section{Ring geometry}

\label{sec:Geometry}

In principle, the effective interactions $C_k$ can be freely altered by using optical tweezers to place the molecules at different distances $R_k$ and angles $\theta_k$ relative to the atom. Additionally, the setup offers a way to modify interactions for all molecules simultaneously, without having to move them. We recall the quantization axis $\vec{e}_0$ is dictated by the external (electric or magnetic) field, \emph{i.e.}, the pseudospin states $\Uparrow,\Downarrow$ ($\uparrow,\downarrow$) are defined as states with specific values of $m_j$ ($M_F$) in reference to the field axis. Therefore, changing the angle of the field in the laboratory frame (before initializing particles into the pseudospin states) is seen, in the frame of each particle, as rotating the entire system while keeping the direction of $\vec{e}_0$ unchanged. This results in changing the polar angles $\theta_k$ for all the molecules simultaneously.

To study the possibilities offered by this technique, in this and following chapters we focus on one particular particle layout: molecules arranged in a ring shape, with the atom at the center. This arrangement allows us to tune the system between a homogeneous case, where all the couplings $C_k$ are equal, and an inhomogeneous case with inequal $C_k$, with the degree of inhomogeneity being tunable.

\begin{figure}[t]
 \centering
 \includegraphics[width=0.5\textwidth]{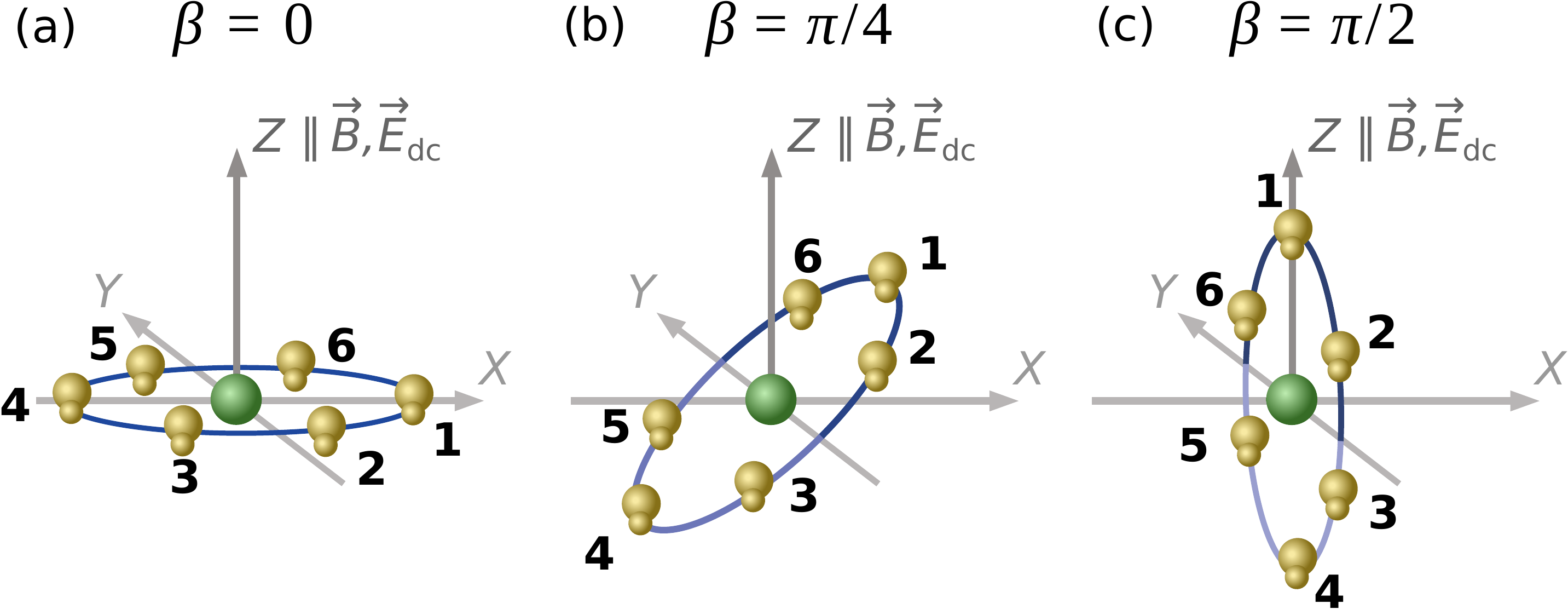}
 \caption{Schematic depiction of an $N_\mathrm{bath}=6$ ring-shaped system at different ring tilt angles $\beta$. The numbering of molecules from $1$ to $N_\mathrm{bath}$ is shown. The $Z$ axis is defined as parallel to the external electric or magnetic field (and to the quantization axis $\vec{e}_0$), while the $X,Y$ axes are defined arbitrarily. The ring can therefore be effectively tilted either by repositioning the molecules, or by changing the angle of the field. The molecule positions are defined by Eq.~\eqref{eq:particle_positions}.}
 \label{fig:diagram_TiltedRing_Clarified}
\end{figure}

The molecules are arranged evenly on the circumference of a ring with radius $r_0$, which we set to $1.5\,\mathrm{\mu m}$. We define the coordinates $X,Y,Z$ so that $Z$ points along the external field, and $X,Y$ are chosen arbitrarily. The position of molecule $k = 1,\ldots,N_\mathrm{bath}$ relative to the atom, $\vec{R}_k = \left( X_k, Y_k, Z_k \right)$, is defined as
\begin{align}
\label{eq:particle_positions}
 X_k &= r_0 \cos \left( \frac{2\pi (k-1)}{N_\mathrm{bath}}\right) \cos \left(\beta\right), \\
 Y_k &= -r_0 \sin \left(\frac{2\pi (k-1)}{N_\mathrm{bath}}\right), \nonumber \\
 Z_k &= r_0 \cos \left(\frac{2\pi (k-1)}{N_\mathrm{bath}}\right) \sin \left(\beta\right). \nonumber
\end{align}
The parameter $\beta$ describes the tilt angle between the ring plane and the X-Y plane, or, equivalently, the angle between the external field and the normal of the ring. Therefore $\beta$ can be changed simply by rotating the external field. The arrangement of particles at different $\beta$ is schematically depicted in Fig.~\ref{fig:diagram_TiltedRing_Clarified}. When $\beta = 0\pi$, the ring lies in the X-Y plane [see Fig.~\ref{fig:diagram_TiltedRing_Clarified}(a)], and for increasing $\beta$, it is tilted counterclockwise around the Y axis, until it fully lies in the Y-Z plane for $\beta = \pi/2$ [see Figs.~\ref{fig:diagram_TiltedRing_Clarified}(b) and~\ref{fig:diagram_TiltedRing_Clarified}(c)]. The resulting polar angles $\theta_k$, relative to the axis $\vec{e}_0 \equiv \vec{e}_Z$, can be calculated as
\begin{equation}
\label{eq:ring-geometry-thetaK}
 \theta_k = \arccos \frac{Z_k}{r_0} = \arccos \left[\cos \left(\frac{2\pi (k-1)}{N}\right) \sin \left(\beta\right)\right].
\end{equation}

\begin{figure}[t]
 \centering
 \includegraphics[width=0.5\textwidth]{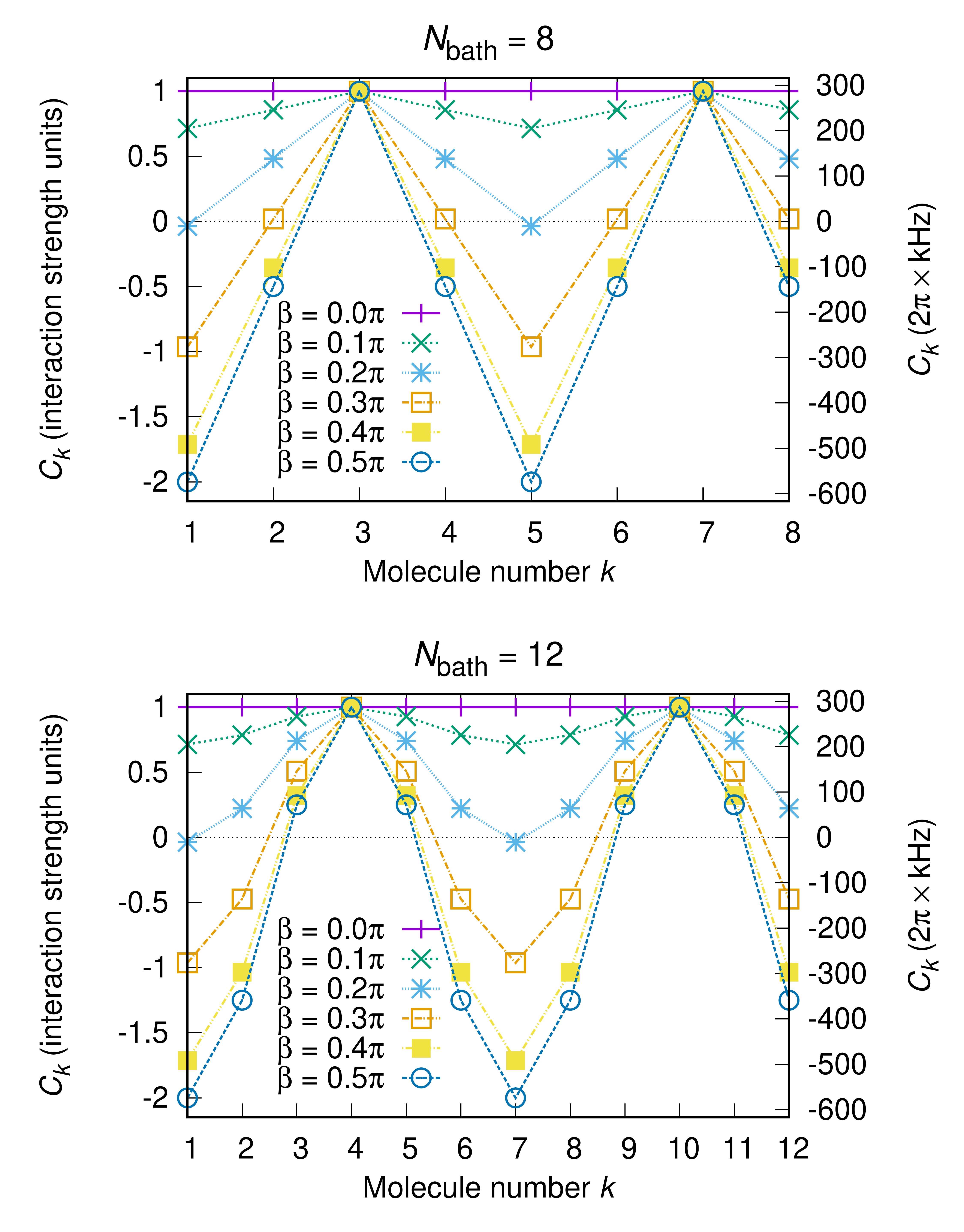}
 \caption{Values of the effective spin-spin interaction coefficients $C_k$ for molecules $k=1,\ldots,N_\mathrm{bath}$ which are distributed on a ring with varying tilt angle $\beta$. We show examples for $N_\mathrm{bath}=8$ and $N_\mathrm{bath}=12$. The values of $C_k$ are calculated from Eq.~\eqref{eq:example_Ck_expression}, as a function of polar angles $\theta_k$ given by Eq.~\eqref{eq:ring-geometry-thetaK}. The lines connecting the points are only to guide the eye. The horizontal dashed line indicates the $C_k = 0$ point. $C_k$ is given both in units of the interaction strength $\mu_\updownarrow \mu_\Updownarrow / (4 \pi \epsilon_0 |R_k|^3)$ (left axis), and in units of $2\pi \times \mathrm{kHz}$ (right axis).}
 \label{fig:LinearCombinationUpspin_Couplings_molnumber}
\end{figure}

Note that the number of molecules on the ring is restricted by realistic limitations on positioning optical tweezers. Assuming that two molecules can be placed no closer than $\approx 700\,\mathrm{nm}$ apart, a ring with a radius of $1.5\,\mathrm{\mu m}$ can accommodate up to about $13$ molecules. Increasing the ring radius can accommodate more molecules, but at the cost of weakening the atom-molecule interactions.

The angle $\beta$ affects the degree of inhomogeneity of the interactions. Figure~\ref{fig:LinearCombinationUpspin_Couplings_molnumber} shows how the obtained effective spin-spin couplings $C_1, C_2, \ldots , C_{N_\mathrm{bath}}$ vary with $\beta$, assuming that the couplings are given by Eq.~\eqref{eq:example_Ck_expression} as in our example system. In the figure $C_k$ is given in two different units: in generic units of the interaction strength $\mu_\updownarrow \mu_\Updownarrow / (4 \pi \epsilon_0 |R_k|^3)$, and in kHz (assuming that $\mu_\updownarrow \mu_\Updownarrow / (4 \pi \epsilon_0 |R_k|^3) = 2\pi \times 287 \,\mathrm{kHz}$ as listed in Table~\ref{tab:CaFKparameters}). At $\beta=0$, all $\theta_k$ are equal and therefore all $C_k$ are identical. As $\beta$ increases, different spins lie at different $\theta_k$, and $C_k$ become increasingly inhomogeneous.

It is worth noting that, for any $\beta$, there exist groups of molecules which share the same value of $\cos^2 \theta_k$, and thus represent bath spins with identical values of $C_k$. For example, for $N_\mathrm{bath}=8$, the molecules numbered $1$ and $5$ (on opposite sides of the ring) share the same value of $C_k$; the same is true for the two molecules $3,7$, and for the four molecules $2,4,6,8$. Such groups appear for both even and odd $N_\mathrm{bath}$, if particle positions are given by Eq.~\eqref{eq:particle_positions}. However, if the ring is rotated around its central axis by some arbitrary angle (corresponding to replacing $2\pi(k-1)$ with $2\pi(k-1) + \mathrm{const}$ in Eq.~\eqref{eq:particle_positions}), this property remains only for even $N_\mathrm{bath}$.

The properties of the central spin system can significantly vary, depending on how inhomogeneous the interactions are. In the next section we will show concrete examples of how modifying the inhomogeneity via the parameter $\beta$ changes the dynamical properties of the system.

\section{Example dynamics}

\label{sec:Dynamics}

  In this section, we present numerical simulations of the time evolution of the described ring system.
  We focus on scenarios that demonstrate how tuning the parameter $\beta$ allows to smoothly tune the simulated system parameters.

	We consider two scenarios.
	In the first scenario, the central spin corresponds to a qubit interacting with a disordered environment, and we examine the ${ \langle \hat{S}^{(0)}_z(t) \rangle }$ dynamics. Modifying $\beta$ is equivalent to changing the disorder level of the simulated environment.
	In the second scenario, all the spins represent a network of qubits, with one central node and $N_\mathrm{bath}$ edge nodes (bath spins). We show that a classical bit can be communicated between two edge nodes that have the same value of $C_k$. The time of the communication process depends on $\beta$.

	Unless stated otherwise, throughout this section we use the system Hamiltonian in Eq.~\eqref{eq:eff-spin-hamiltonian}, the values of $C_k$ from Eq.~\eqref{eq:example_Ck_expression}, the angles $\theta_k$ from Eq.~\eqref{eq:ring-geometry-thetaK}, and the physical parameters $c_S,c_0,c_\Delta$ listed in Table~\ref{tab:CaFKparameters}.

\subsection{Central spin decoherence}

\label{sec:Dynamics-Decoherence}

The case of a qubit (central spin) coupled to a disordered environment (spin bath) is a common scenario in which inhomogeneity of $C_k$ can have significant impact on system properties.
	The qubit starts in a pure quantum state, but over time, interactions generate quantum correlations between the qubit and environment, leading to qubit decoherence.
	The decoherence process manifests as a decay of the norm of the central spin vector, $||\langle \hat{\vec{S}}^{(0)} \rangle|| \equiv \sqrt{\langle \hat{S}^{(0)}_+ \rangle \langle \hat{S}^{(0)}_- \rangle + \langle \hat{S}^{(0)}_z \rangle^2 }$; this can be seen directly by analyzing the reduced density matrix of the central spin~\cite{2002-Schliemann-PRB,2016-Yang-RepProgPhys}.
	For this reason, dynamics of the central spin are of interest from a quantum information point of view.
	In a particularly simple case, the initial state has a single, well-defined value of total spin $z$ projection ($\langle \hat{\sigma}_z \rangle$) and so ${ \langle \hat{S}^{(0)}_\pm \rangle(t)=0 }$ at all times. Then the decoherence is fully captured by the decay of ${ \langle \hat{S}^{(0)}_z \rangle }$. In further discussion, we will focus on this kind of initial states.

	A commonly considered example of such a scenario is that of an electron qubit in a semiconductor quantum dot, which interacts with neighboring nuclear spins through a hyperfine contact interaction.
	In that case, the interaction strengths for different nuclei are inhomogeneous because of the spatial variation of the electron wave function.
	The hyperfine interactions are realistically modeled by isotropic $XXX$ couplings, of the form ${ \hat{\vec{S}}^{(0)} \cdot \hat{\vec{S}}^{(k)} } = { \hat{S}^{(0)}_z \hat{S}^{(k)}_z + \frac{1}{2}( \hat{S}^{(0)}_+ \hat{S}^{(k)}_- + \mathrm{H.c.}) }$.
	For this reason, spin dynamics in such systems have been primarily studied in the $XXX$ central spin model~\cite{2002-Khaetskii-PRL,2002-Schliemann-PRB,2003-Khaetskii-PRB,2007-Zhang-JPhysCondMat,2007-Bortz-JStatMech,2007-Bortz-PRB,2010-Bortz-PRB,2020-Wan-QuantInfProc}.
	These studies reveal that central spin dynamics are significantly different, depending on whether the interactions are homogeneous or not.
	For a homogeneously interacting $XXX$ system (and vanishing Zeeman fields $c_S,c_0 = 0$), the dynamics are exactly periodic~\cite{2007-Bortz-JStatMech}.
	The central spin $z$ initially decays on short timescales, but then returns to the initial state, and oscillates in this way indefinitely.
	Therefore, the initial pure state is never lost permanently, and the system features no long-term decoherence.
	For inhomogeneous couplings, the behavior is different.
	Initially, the central spin $z$ oscillates periodically, but over time the oscillation envelope decays. Eventually, the central spin settles at a diminished value~\cite{2002-Schliemann-PRB}.
	This long-time spin value depends on the initial bath polarization; in particular, for a bath with initial zero total polarization, the central spin decays nearly completely to zero.
	The timescale of this decay also varies, depending on the initial bath state, and on the central-bath coupling coefficients. Usually, it is inversely proportional to the spread (difference between maximal and minimal values) of the coupling coefficients~\cite{2002-Khaetskii-PRL,2003-Khaetskii-PRB,2007-Bortz-PRB,2010-Bortz-PRB}.

Our setup simulates a different scenario where environmental couplings are solely of the $XX$ kind, and to our knowledge, the decay dynamics in this model have not yet been studied in detail.
	Nevertheless, the dynamics of the $XX$ and $XXX$ models are qualitatively similar in the most important aspects (though they are not identical), and in particular the $XX$ model displays similar decoherence dynamics. Since the ring tilt parameter $\beta$ allows us to smoothly set the inhomogeneity of $C_k$ (\emph{i.e.}, the ``disorder level'' of the simulated environment), it means our setup can be used directly to explore this decoherence in $XX$ systems in homogeneous vs. inhomogeneous coupling cases.

\begin{figure}[t]
 \centering
 \includegraphics[width=0.5\textwidth]{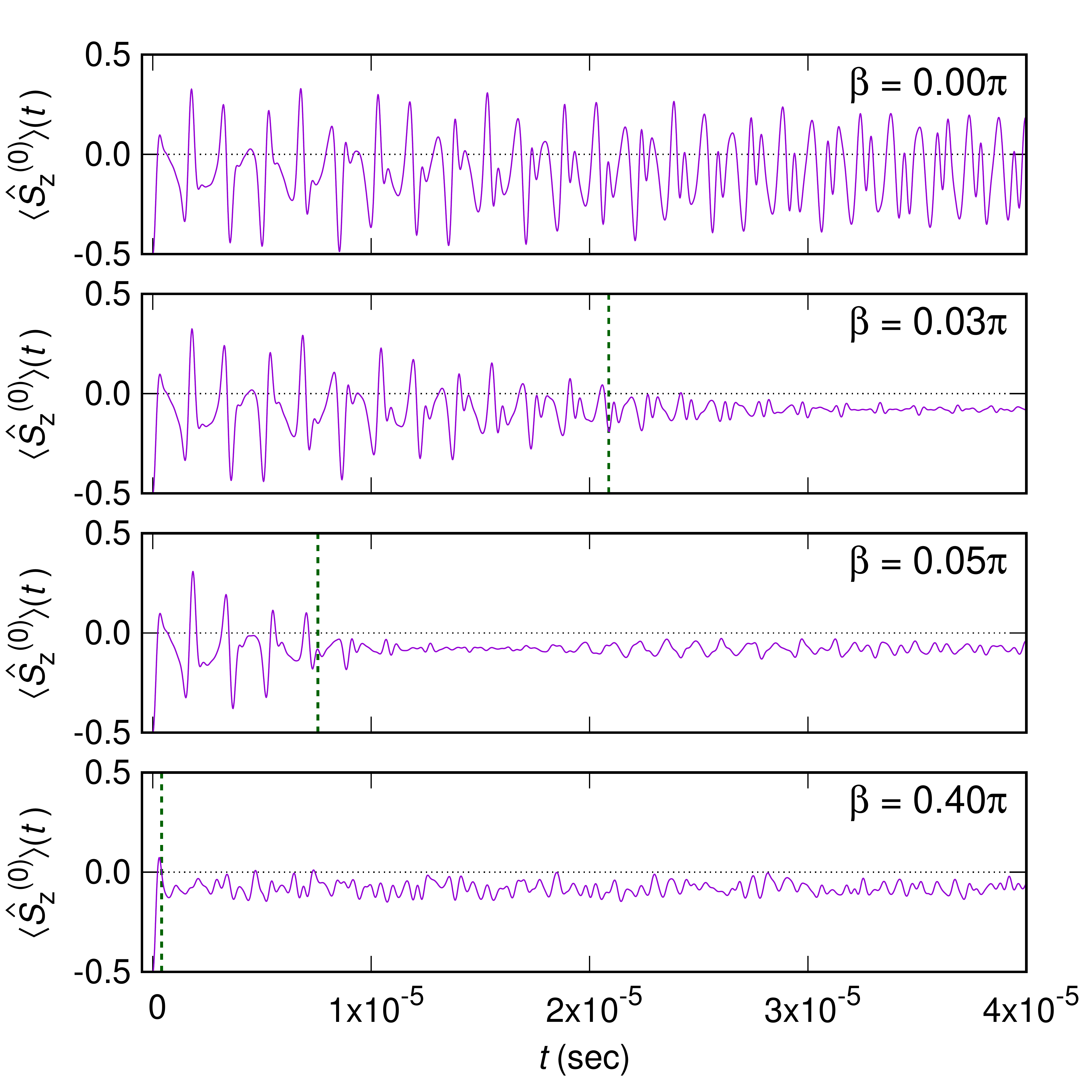}
 \caption{Time evolution of the central spin $z$-projection ${ \langle \hat{S}^{(0)}_z \rangle (t) }$ for different values of $\beta$, for an example $N_\mathrm{bath}=12$ system with the initial state given by Eq.~\eqref{eq:random-initial-state}. At $\beta = 0$, all couplings are homogeneous and the central spin does not exhibit significant long-time decay, but displays large-amplitude oscillations. For larger $\beta$, the initial oscillations decay after a certain time, stabilizing as small oscillations around a diminished value. The vertical dashed lines indicate the predicted decay timescales $\tau = 1/A$ at given $\beta > 0$ [Eq.~\eqref{eq:decay-timescale}].}
 \label{fig:SpinEvolution_HalfBathDecayN12-4rows-decaytime}
\end{figure}

We set the system size to $N_\mathrm{bath}=12$, and take the initial state
	\begin{equation}
	\label{eq:random-initial-state}
	 |\Psi_0\rangle = \sum_{\{ |j\rangle \}} \alpha_j |\Downarrow\rangle \otimes |j\rangle.
	\end{equation}
	The sum includes all the possible bath configurations $|j\rangle \equiv { |S^{(1)}, \ldots, S^{(N_\mathrm{bath})} \rangle }$ that have six bath spins pointing up and six bath spins pointing down.
	The coefficients $\alpha_j$ have magnitudes and phases drawn independently from a uniform random distribution, and are normalized to $\sum |\alpha_j|^2 = 1$.
	This initial state represents a polarized qubit in a bath with zero overall polarization.

	The evolution of ${ \langle \hat{S}^{(0)}_z \rangle(t) }$ is shown in Fig.~\ref{fig:SpinEvolution_HalfBathDecayN12-4rows-decaytime} at several different values of $\beta$.
	When $\beta = 0$ (first row), all couplings are identical. Unlike the $XXX$ model, the $XX$ model dynamics in this case are not periodic, but ${ \langle \hat{S}^{(0)}_z \rangle (t) }$ still displays high-amplitude oscillations with no significant long-time decay, repeatedly returning close to its initial value. From the decoherence point of view, although the qubit becomes entangled with the environment, it keeps returning to an almost-pure state.

	For $\beta > 0$, the dynamics are different.
	At short times, the spin oscillates identically with the $\beta = 0$ limit, but now the envelope of this oscillation decays over time.
	Eventually, the central spin settles around a near-zero value with only small-amplitude oscillations. This behavior represents irreversible decoherence, \emph{i.e.}, the qubit does not return to a pure state on reasonable timescales.
	The time needed for the oscillation envelope to decay to its final value, which we designate $\tau$, is dependent on $\beta$: as $\beta$ increases and the interactions become more inhomogeneous, the decay becomes faster. In these aspects the dynamics are qualitatively highly similar to the $XXX$ model.

	We can attempt to more precisely describe the relationship between $\beta$ and $\tau$.
	To our knowledge, the decay timescale for the $XX$ model has not been analytically derived yet.
	Therefore, we estimate it by assuming that, like in the $XXX$ model, it is inversely proportional to the spread of couplings $A = \max(|C_k|) - \min(|C_k|)$:
	\begin{equation}
	\label{eq:decay-timescale}
	 \tau = 1/A.
	\end{equation}
	Note that $\tau \rightarrow \infty$ for fully homogeneous couplings, which agrees with the observation that no long-time decay occurs for $\beta = 0$.

	\begin{figure}[t]
 \centering
 \includegraphics[width=0.5\textwidth]{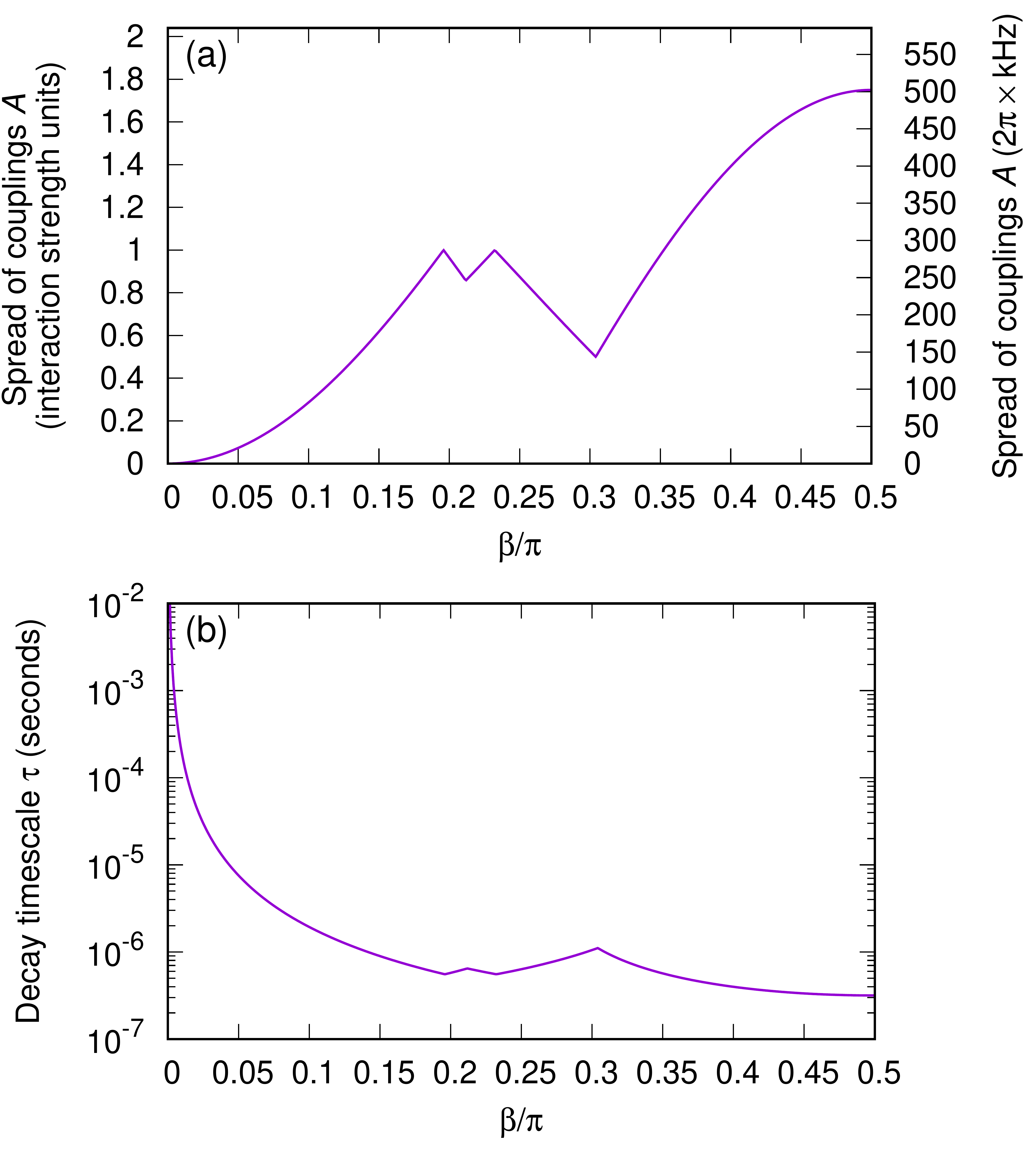}
 \caption{(a) The spread of couplings ${ A = \max(|C_k|) - \min(|C_k|) }$, as a function of the tilt angle $\beta$, for the example $^{39}$K--$^{40}$Ca$^{19}$F system with $N_\mathrm{bath}=12$. Several kinks are visible in the vicinity of $\beta \approx 0.2\pi$ and $\beta \approx 0.3\pi$; see text for explanation. Similarly as in Fig.~\ref{fig:LinearCombinationUpspin_Couplings_molnumber}, $A$ is given both in units of $\mu_{01} \mu_\Updownarrow / (4 \pi \epsilon_0 |R_k|^3)$ and in units of $2\pi \times\,\mathrm{kHz}$. (b) The resulting predicted timescale for the central spin decay, $\tau = 1/A$, in units of seconds.}
 \label{fig:LinearCombinationUpspin_Amplitude_Timescale}
\end{figure}

	To check the accuracy of this estimate, in Fig.~\ref{fig:SpinEvolution_HalfBathDecayN12-4rows-decaytime} we have indicated the calculated values of $\tau$ at given $\beta$ with a green dashed line.
	Interestingly, despite the crudity of the estimation, the figure shows that Eq.~\eqref{eq:decay-timescale} provides a remarkably close approximation of the decay time.

	To directly show the relationship between $\tau$ and $\beta$, we show the spread $A$ as a function of $\beta$ in Fig.~\ref{fig:LinearCombinationUpspin_Amplitude_Timescale}(a), and the corresponding timescale $\tau = 1/A$ in Fig.~\ref{fig:LinearCombinationUpspin_Amplitude_Timescale}(b), for the example $N_\mathrm{bath}=12$ system.
	For most values of $\beta$, $A(\beta)$ increases monotonically [and $\tau(\beta)$ consequently decreases], except for the region $0.2\pi < \beta < 0.3\pi$ where there are several kinks in both plots.
	They appear because $A$ is defined as the difference between the \emph{absolute values} $|C_k|$, which are not necessarily smooth functions of $\beta$.
	Overall, however, regulating $\beta$ allows us to smoothly tune the value of $\tau$ across several orders of magnitude.

Finally, we briefly describe additional simulations (not shown) which we have done to check the results of adjusting different system parameters.

First, we have investigated the results of adjusting the parameter $c_\Delta$, which represents an effective Zeeman field acting on the central spin, and is controllable via external fields. So far throughout this section, we have considered a system with a Zeeman field strength $c_\Delta = 2\pi \times 19 \,\mathrm{kHz}$, which is small compared with the interaction energies.
	We have additionally checked the dynamics for the same system but with $c_\Delta$ set to zero, and found that the dynamics are almost unchanged, confirming that such small $c_\Delta$ has a negligible effect on the dynamics.
	On the other hand, if $c_\Delta$ becomes large compared with interaction strengths, it might have a significant effect.
	We have therefore checked the central spin dynamics for $N_\mathrm{bath}=12$ and $\beta = 0.10\pi$, with all system parameters set as previously, but with an artificially increased mismatch $c_\Delta$ to represent off-resonance between the qubit and the environment.
	As $c_\Delta$ becomes large in comparison to interaction strengths, we find that spin exchange between the central and bath spins becomes suppressed because of its increasing energy cost, and only a small part of the central spin decays at long times.
	This behavior is similar to $XXX$ model dynamics, where a large Zeeman field results in only a small part of the initial spin decaying~\cite{2003-Khaetskii-PRB}.

Second, we have checked the dynamics for different bath polarizations, \emph{i.e.}, when setting initial states similar to Eq.~\eqref{eq:random-initial-state} but with a smaller number of up-pointing spins in the bath. We find that, for increasing polarization, a smaller part of the initial spin decays, which is again similar to the $XXX$ model results~\cite{2002-Schliemann-PRB}. Furthermore, we find that at increasing bath polarization, the decay timescale in our model increases so that $\tau$ is no longer a good estimate.

Third, we have compared $XXX$ and $XX$ dynamics for the unpolarized bath initial state [Eq.~\eqref{eq:random-initial-state}] in order to confirm that they are indeed similar. Specifically, we have compared dynamics for two Hamiltonians: one with $XX$ couplings, $c_\Delta \hat{S}^{(0)}_z + \sum_k |C_k| (\hat{S}^{(0)}_+ \hat{S}_- + \mathrm{H.c.})$, and one with $XXX$ couplings, $c_\Delta \hat{S}^{(0)}_z + \sum_k |C_k| (\hat{S}^{(0)}_+ \hat{S}_- + \hat{S}^{(0)}_- \hat{S}_+ + 2 \hat{S}^{(0)}_z \hat{S}^{(k)}_z)$. For the purposes of this comparison, we explicitly set all $C_k$ to be real and positive, the same way they are in the previously cited $XXX$ model papers (note that phases of $C_k$ in our setup can be modified; see Appendix~\ref{sec:Phase-Transformation}).
	Our simulations show that the dynamics of ${ \langle \hat{S}^{(0)}_z \rangle(t) }$ remain quantitatively similar between both models, with approximately the same decay timescale and long-time average of ${ \langle \hat{S}^{(0)}_z \rangle(t) }$. This suggests that our setup can be used to approximately study $XXX$ dynamics in certain specific cases.
	However, this only applies to the case with unpolarized bath and small $c_\Delta$.
	When simulations are run with higher bath polarizations or stronger Zeeman fields $c_\Delta$, we find the $XX$ and $XXX$ model dynamics deviate from each other.

\subsection{Quantum network simulation}

\label{sec:Dynamics-Networks}

The central spin Hamiltonian can also describe a star-shaped quantum network, with several edge qubits coupled to a single central qubit. In this case, instead of regarding the bath spins as a collective environment, we are interested in the individual time evolution of each bath spin. (Although the designation ``bath spins'' is no longer fully appropriate in this case, for consistency we will keep using it throughout this section.) By adjusting the physical parameter $\beta$, the couplings within the network can be changed.

One example of a quantum network application, where the inhomogeneity of interactions directly affects the outcome, is the protocol of generating entangled states described in Ref.~\cite{2008-Ferraro-EPJSpecTop}. In this procedure, a star-shaped quantum network with homogeneous $XX$ couplings is initialized in a state ${ |\Uparrow ; \downarrow^{(1)}, \ldots, \downarrow^{(N_\mathrm{bath})}\rangle }$, then measurements of the central spin state are performed at specific evolution times. If the measurement outcome is $\Downarrow$, the bath spins are in a state ${ \propto \sum_k C_k |\downarrow\ldots \uparrow^{(k)} \ldots\downarrow\rangle }$. In the homogeneous case (corresponding to $\beta= 0$), the obtained bath state is a W-state where each edge spin has equal probability of pointing up. For inhomogeneous $C_k$, generalized W-like states are instead obtained. The protocol can be directly realized in our setup, and $\beta$ then serves as a control parameter for modifying the form of entanglements created within the network.

Another interesting application of quantum networks is so-called quantum state transfer, \emph{i.e.}, transmitting an arbitrary quantum state from one part of the network to another. Typically, an input bath qubit is initialized in some desired pure quantum state, while the rest of the system is in some neutral state. Then the system is allowed to evolve, and, after some (predictable) optimal transfer time $\tau_\mathrm{transf}$, this quantum state is transferred with a high fidelity to an output bath spin. Quantum state transfer has been studied for quantum networks of different shapes~\cite{2003-Bose-PRL,2007-Bose-ContempPhys}, including spin stars with $XX$ couplings~\cite{2005-DeChiara-PRA,2007-Jiang-PRA,2008-HongLiang-JPB,2011-Yung-JPB,2013-Salimi-QIP}.

In principle, it is possible to simulate such a transfer protocol with our setup, but in this case we run into an experimental limitation. In typical transfer protocols, it is assumed that effective Zeeman fields acting on the spins are no stronger than the interactions. On the other hand, in our setup $c_0,c_S$ are locked to values much larger than the interaction. This limitation, as noted in Section~\ref{sec:Limitations}, makes it difficult to realize a transfer protocol that faithfully preserves the phase between the input state components. This is a limitation, since the accurate transfer of an arbitrary qubit state requires preserving both the magnitude and phase of its components.

In light of this limitation, we shift our focus to a simpler task of transmitting a classical bit (i.e., the input spin is either in state $|\uparrow\rangle$ or $|\downarrow\rangle$), rather than a full qubit state. In that case, the limitation mentioned above does not pose a
problem, as there is no phase information to transmit. Note that the idea of transmitting classical information through a spin network has
been considered before~\cite{2004-Yung-QIC}. The protocol we will now describe allows us to transmit a single bit between two edge spins which share identical values of $C_k$; this makes the ring system especially well-suited for its study, since pairs of such edge spins appear naturally in the ring geometry. While less exciting from a quantum information standpoint, this procedure might be a foundation for a usable qubit state transfer protocol if the fields $c_0$, $c_S$ could be controlled.

We present qualitative numerical results to demonstrate the feasibility and properties of this specific protocol. Our focus here is on presenting the basic properties of this scenario, and the impact of $\beta$ on system properties, therefore we have not attempted detailed theoretical analysis, and we do not attempt to exactly quantify the efficiency of the transfer.

\begin{figure}[t]
 \centering
 \includegraphics[width=0.50\textwidth]{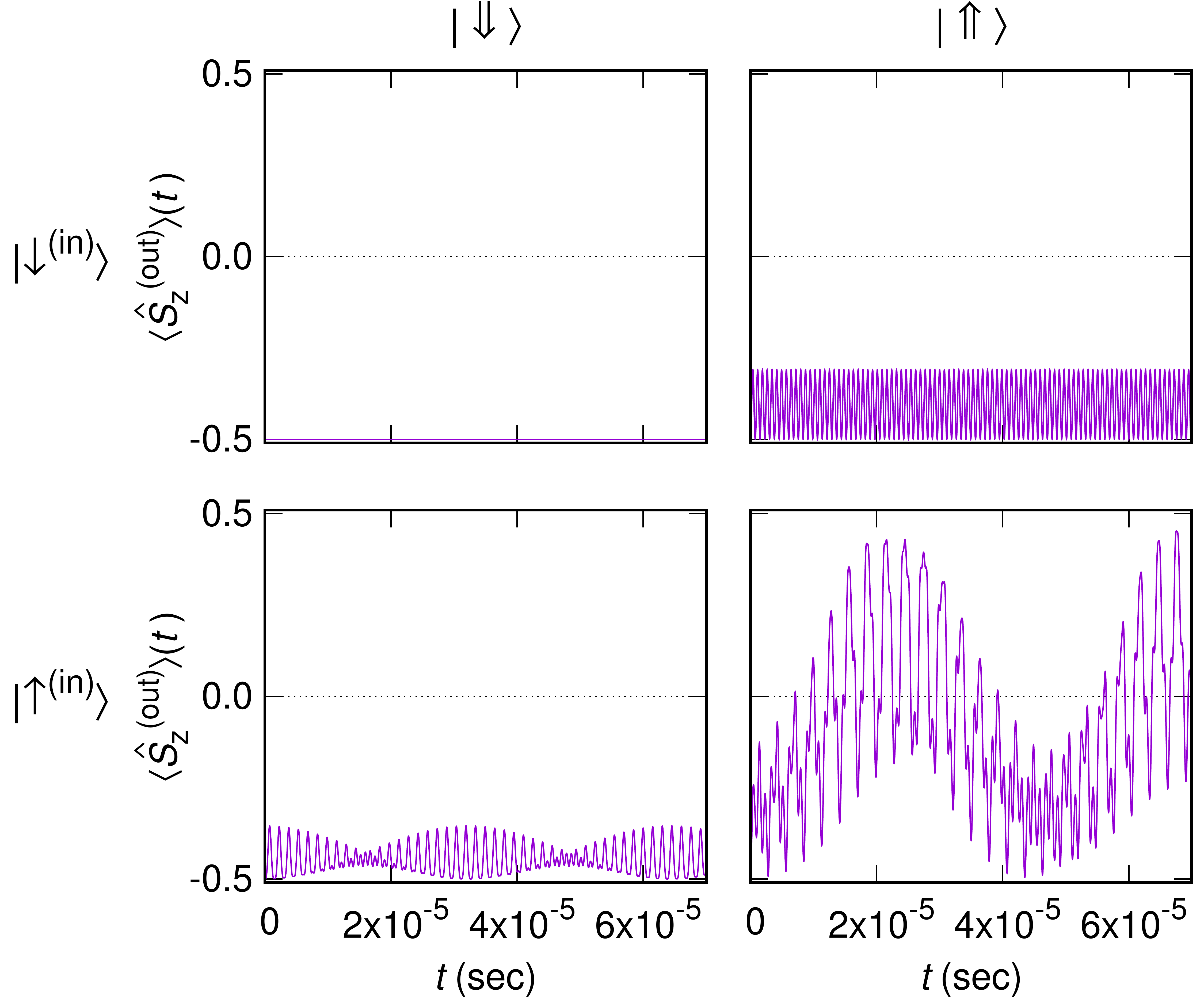}
 \caption{Time evolution of the $z$-component of the output bath spin $\langle\hat{S}_z^{(\mathrm{out})}\rangle$ in an $N_\mathrm{bath}=8$ system. The initial state is ${ |S^{(0)} S^{(\mathrm{in})} \downarrow_2\ldots \downarrow_8\rangle }$, with $S^{(0)} = \Uparrow$ or $\Downarrow$, $S^{(\mathrm{in})} = \uparrow$ or $\downarrow$. Here the input and output spins are chosen as $S^{(\mathrm{in})} = S^{(\mathrm{3})}$ and $S^{(\mathrm{out})} = S^{(\mathrm{7})}$. The system is arranged in a ring geometry with angle $\beta = 0.12\pi$, so that $C_3 = C_7 \ne$ other $C_k$. The different plots correspond to different initial states of $S^{(\mathrm{in})}$ and $S^{(0)}$, as indicated by the the row and column labels. For the initial state with $|\Uparrow \uparrow^\mathrm{(\mathrm{in})}\rangle$ (bottom right), the output spin $z$ gradually evolves from $-1/2$ to $+1/2$, reaching a maximum in a time $\tau_\mathrm{transf} \approx 2 \times 10^{-5}\,\mathrm{s}$. For all other initial states, the output spin $z$ remains close to $-1/2$ at all times. Effectively, if the initial central spin state is $\Uparrow$ (right column), the initial state of the input spin is transmitted within time $\tau_\mathrm{transf}$ to the output spin.}
 \label{fig:QST_MultipleIniStates_2x2}
\end{figure}

We begin by selecting two bath spins as the input and the output spin, with the condition that their values of $C_k$ are identical ($C_\mathrm{in} = C_\mathrm{out}$). For the following demonstration, we take an $N_\mathrm{bath} = 8$ system, and set spins 3 and 7 as input and output, respectively. In this case, the condition $C_3 = C_7$ is fulfilled at any $\beta$. Additionally, note that $C_3, C_7$ are both independent of $\beta$ (see Fig.~\ref{fig:LinearCombinationUpspin_Couplings_molnumber} in Section~\ref{sec:Geometry}). This simplifies matters, because it guarantees that these interactions will not become zero at any value of $\beta$. The system is prepared in an initial state of the form ${ |S^{(0)}; S^{(\mathrm{in})}, \downarrow, \ldots ,\downarrow\rangle }$, where the central spin state $S^{(0)}$ is $\Uparrow$ or $\Downarrow$, the input spin state $S^{(\mathrm{in})}$ is $\uparrow$ or $\downarrow$, and all other bath spins (including the output spin) are in state $\downarrow$.

In Fig.~\ref{fig:QST_MultipleIniStates_2x2} we present the dynamics of the output spin $\langle \hat{S}^{(\mathrm{out})}_z \rangle(t)$ in an example system with $\beta = 0.12\pi$. This evolution is shown for each of the four possible initial states of the input and central spin (as indicated by the row and column labels).

For the initial states ${ |\Downarrow \downarrow^{(\mathrm{in})} \rangle,} { | \Downarrow \uparrow^{(\mathrm{in})}  \rangle, } { | \Uparrow  \downarrow^{(\mathrm{in})} \rangle }$, the value of $\langle \hat{S}^{(\mathrm{out})}_z \rangle$ remains almost unchanged from its initial value, remaining close to $-1/2$ at all times. However, for the initial state with ${ |\Uparrow \uparrow^{(\mathrm{in})} \rangle }$, the dynamics of $\langle \hat{S}^{(\mathrm{out})}_z \rangle$ are visibly different. $\langle \hat{S}^{(\mathrm{out})}_z \rangle$ oscillates between $-1/2$ and a value very close to $\approx +1/2$. We can define a time $\tau_\mathrm{transf}$ as the time where $\langle \hat{S}^{(\mathrm{out})}_z \rangle$ first reaches the maximum point of the large oscillation. The envelope of this large oscillation is modulated by rapid small-amplitude oscillations.

This result can be summarized as follows. If the central spin is initially set to $\Uparrow$, then within the time $\tau_\mathrm{transf}$ the initial state of the input spin will be transmitted to the output spin, \emph{i.e.}, a measurement of ${ \langle \hat{S}_z^\mathrm{(\mathrm{out})} \rangle (t) }$ will, with near-unity probability, yield a result identical to the initial input spin value. This is equivalent to a communication protocol, where the initial state of the input spin (a single classical bit) can be picked up by measuring the output spin at $\tau_\mathrm{transf}$. On the other hand, if the central spin is set to $\Downarrow$, then no transfer occurs, and ${ \langle \hat{S}_z^{(\mathrm{out})}\rangle(t) }$ remains near $-1/2$. In addition to transmitting a bit, the procedure can be also interpreted as implementing a classical \textsc{AND} operation: if the up-spin (down-spin) states are mapped to 1 and 0 respectively, then ${ \langle \hat{S}^{(\mathrm{out})}\rangle (\tau_\mathrm{transf}) }$ is approximately equal to the result of performing the \textsc{AND} operation on the initial choices for $S^{(0)},S^{(\mathrm{in})}$.

\begin{figure}[t!]
 \centering
 \includegraphics[width=0.50\textwidth]{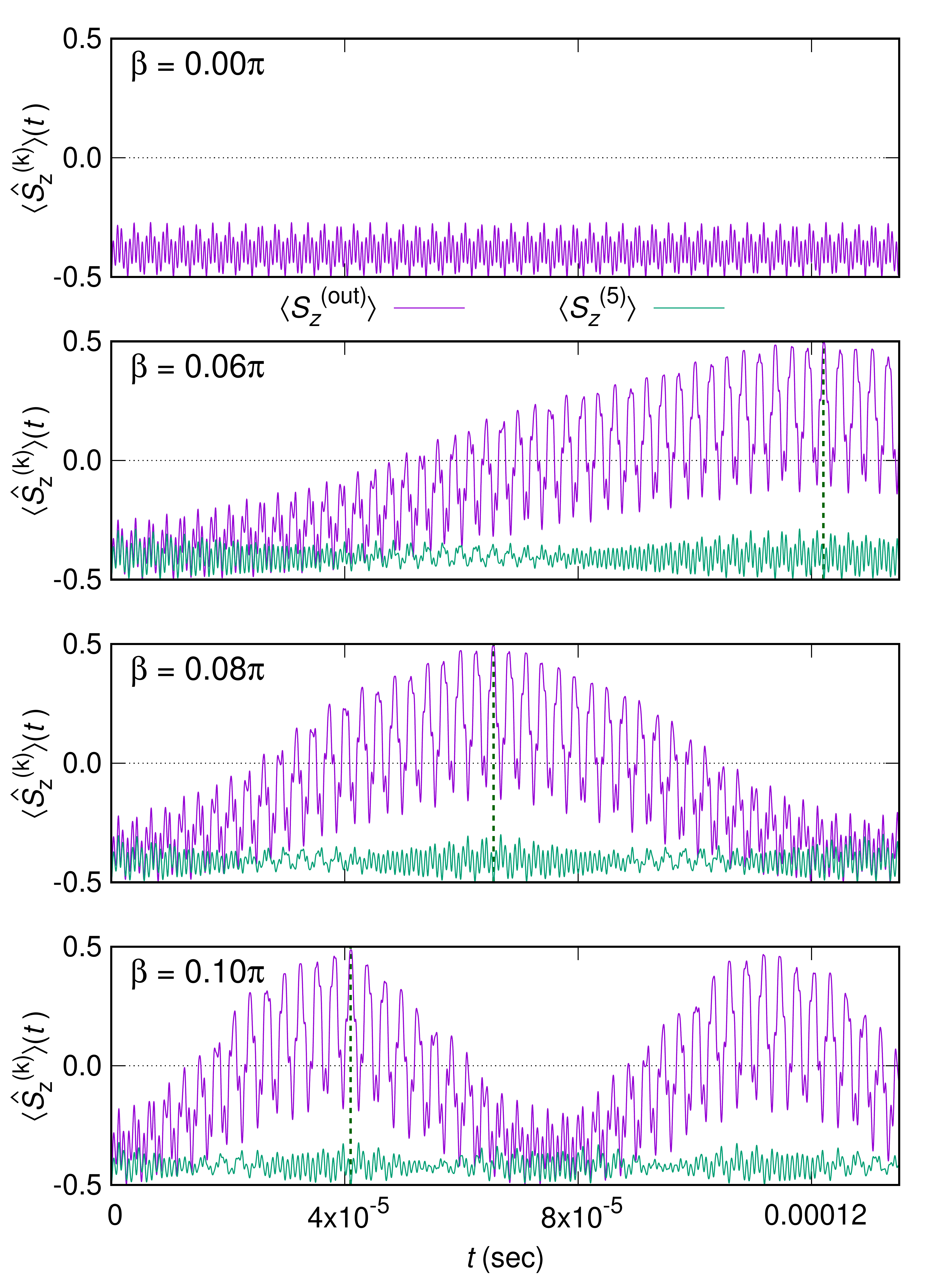}
 \caption{Time evolution of the bath spin $\langle \hat{S}^{(\mathrm{out})}_z \rangle(t) \equiv \langle \hat{S}^{(7)}_z \rangle (t)$ for an initial state ${ |\Uparrow \uparrow_1 \downarrow_2 \ldots \downarrow_N\rangle }$, with same system parameters as in Fig.~\ref{fig:QST_MultipleIniStates_2x2}, but with changing $\beta$. The transfer period $\tau_\mathrm{transf}$, defined as the time where $\langle \hat{S}^{(\mathrm{out})}_z \rangle (t)$ first reaches its maximum value, is marked for $\beta > 0$ with a dashed line. Additionally, for $\beta > 0$, the evolution of spin $\langle \hat{S}^{(5)}_z \rangle$ is shown. As $\beta$ becomes larger (the differences between $C_3 = C_7$ and other $C_k$ become larger), the period of the transmission from spin 3 to 7 becomes smaller. Regardless of $\beta$, the transmission does not occur for other bath spins (such as spin 5) that have $C_k$ different from $C_3 = C_7$, as can be seen from evolution of $\langle \hat{S}^{(5)}_z \rangle$.}
 \label{fig:QST_BathToBath_N8}
\end{figure}

The transfer period $\tau_\mathrm{transf}$ can be regulated via the parameter $\beta$. Figure~\ref{fig:QST_BathToBath_N8} shows the dynamics of $\langle \hat{S}^{(\mathrm{out})}_z \rangle (t)$ for the initial state $| \Uparrow \uparrow^\mathrm{(\mathrm{in})} \rangle$, for increasing values of $\beta$. The dashed green lines indicate $\tau_\mathrm{transf}$, defined as the time where $\langle \hat{S}^{(\mathrm{out})}_z \rangle (t)$ first reaches its maximum value. For $\beta > 0$, we additionally show the evolution of $\langle \hat{S}^{(5)}_z\rangle(t)$, to represent bath spins which have a different value of $C_k$ than $C_\mathrm{in}$.

There are two key observations to be made from the figure. First, for $\beta = 0$ the transfer never occurs (${ \tau_\mathrm{transf} \rightarrow \infty }$). In this limit, all bath spins other than ${ \langle \hat{S}^{\mathrm{(\mathrm{in})}}_z \rangle }$ show identical evolution. As $\beta$ increases, the timescale $\tau_\mathrm{transf}$ becomes shorter. Recalling that, in this example case, the couplings for the input and output spins $C_3 = C_7$ are independent of $\beta$, we conclude the change in $\tau_\mathrm{transf}$ comes from the change in values of all other $C_k$. This indicates that the decrease of $\tau_\mathrm{transf}$ is caused by a growing overall difference between $C_\mathrm{in} = C_\mathrm{out}$ and the other couplings, although we have not attempted to determine an exact form of this relationship.

Second, for bath spins $k$ with $C_k \ne C_\mathrm{in}$, ${ \langle \hat{S}^{(k)}_z \rangle }$ never rises far from the initial value $-1/2$. This confirms that the communication occurs only to the output spin which has the same $C_k$ value as the input spin.

Finally, we mention the effects of changing the parameter $c_\Delta$ while leaving all other Hamiltonian parameters unchanged. We have performed additional numerical simulations, and found that the transmission dynamics in the initial state $|\Uparrow \uparrow^{(\mathrm{in})}\rangle$ are robust to increasing $c_\Delta$. Even at high $c_\Delta \sim 10C_\mathrm{in}$, the dynamics of $\langle \hat{S}^\mathrm{out}_z \rangle$ remain similar, with an oscillation between $-1/2$ and $+1/2$, although the timescale $\tau_\mathrm{transf}$ becomes longer.

\section{Experimental feasibility}

\label{sec:Experimental}

\subsection{Optical tweezer traps}

Let us now analyze the experimental feasibility of realizing the presented setup. One of the most important requirements is the ability to precisely position the particles in space. The uncertainty of the particle positions should be much smaller than the interparticle distances, to avoid significant overlap between particle wave functions and to only consider long-range dipolar interactions.

Such tight trapping of molecules is experimentally feasible, as indicated by existing experiments with molecules and atoms trapped in optical tweezers. For example, in the experiment described in Ref.~\cite{2022-Holland-PRL}, CaF molecules were held in tweezer traps with effective harmonic trapping radial (axial) frequencies of $\omega/(2\pi) = 187.7\,\mathrm{kHz}$ ($42.2\,\mathrm{kHz}$). For a particle of $^{40}$Ca$^{19}$F mass $m = 59\,\mathrm{u}$, this corresponds to radial (axial) position uncertainties $\sqrt{\hbar/m\omega}$ of $0.03\,\mathrm{\mu m}$ ($0.06\,\mathrm{\mu m}$). As an example of similarly strong atom trapping, the experiment in Ref.~\cite{2019-Wang-PRA} featured $^{85}\mathrm{Rb}$ and $^{87}\mathrm{Rb}$ atoms trapped in tweezers with effective radial (axial) trap frequencies $\omega/(2\pi) = 165\,\mathrm{kHz}$ ($27\,\mathrm{kHz}$), corresponding to radial (axial) atom position uncertainties of about $0.03\,\mathrm{\mu m}$ ($0.07\,\mathrm{\mu m}$).

Note that, if harmonic trapping frequencies on the order of ${ 2\pi \times 10^4\,\mathrm{Hz} }$ are used in our setup, the resulting trap excitation energies are small in comparison with the example atom-molecule interaction strength on the order of ${ 2\pi \times 10^5\,\mathrm{Hz} }$. As a result, the dipolar interactions can cause excitations of higher trap states~\cite{2021-Sroczynska-NJP}. If interparticle distances are small in comparison with the extent of the particle wave functions, then the trap excitations can in turn significantly affect the interaction strength, and cannot be kept out of the analysis. However, the distances assumed in our setup are large enough to avoid this effect. To verify this, we have performed simplified numerical calculations to analyze the effect of trap excitations on the atom-molecule interaction in our system. We assumed the atom and molecule are separated by $1.5\,\mathrm{\mu m}$ and placed in ``one-dimensional'' (1D) traps that have infinite radial frequency and a finite axial frequency $\omega = 2\pi \times 50\,\mathrm{kHz}$. The obtained results indicate that the dipole-dipole interaction strength remains almost unchanged, even when particles are excited to the first few excited trap states. This justifies neglecting trap effects in our analysis.

Optical lattices can be used as an alternative to optical tweezer trapping. They are suitable for arranging particles in regular configurations, as long as particle hopping between sites is suppressed. While they do not allow positioning particles with as much flexibility as optical tweezers, lattices can be created in various shapes, such as hexagonal~\cite{2022-Liu-OpticsExpress} or ring-shaped~\cite{2005-Amico-PRL} two-dimensional structures.

\subsection{Rydberg atom trapping}

The necessity to trap the Rydberg atom in a fixed position adds complexity to the setup. Ground-state atoms are typically trapped in red-detuned optical traps, which however become antitrapping for atoms excited to Rydberg states. In most experiments, this issue is addressed by switching off the traps prior to the Rydberg excitation, but this limits the maximum experiment time because of atom expansion. However, recently various methods have been explored to keep an atom trapped even after Rydberg excitation. Examples include bottle potentials made up of combinations of light beams~\cite{2020-Barredo-PRL}, optical lattice potentials which are rapidly inverted simultaneously with the atom's excitation~\cite{2011-Anderson-PRL}, trapping a Rydberg alkaline-earth atom by exploiting its nonexcited ionic core~\cite{2022-Wilson-PRL}, or state-insensitive, magic-wavelength optical lattices that can trap ground and Rydberg states equally~\cite{2003-Safronova-PRA,2013-Li-Nature,2018-Lampen-PRA,2022-Mei-PRL}. The trapping of Rydberg atoms, therefore, appears to be a manageable challenge.

 \subsection{Time limits imposed by molecular state coherence}

The primary limitation to experiment time is the radiative lifetime of Rydberg states (on the order of $10^{-4}\,\mathrm{s}$). Coherence times of molecular states, which are finite in real experiments, in theory can also limit the experiment time. However, in practice they can be made much longer than Rydberg state lifetimes. For example, for a single tweezer-trapped CaF molecule in a combination of $N=0$/$N=1$ states, rotational state coherence times can be on the order of $100\,\mathrm{ms}$~\cite{2021-Burchesky-PRL}. Additionally, coherence between hyperfine states within the same rotational level can last even longer. As an example, for individual nuclear hyperfine $N=0$ states of RbCs, coherence times can exceed $\sim 1\,\mathrm{s}$~\cite{2021-Gregory-NatPhys}. We conclude that molecular state coherence times should not pose a significant limitation.

\section{Conclusion}

\label{sec:Conclusion}

In summary, this paper proposes a design for an ultracold particle system which can act as a quantum simulator for a central spin model. The system consists of a single Rydberg atom acting as the central spin, and polar molecules acting as the bath spins. By making use of electric dipole-dipole interactions between the atom and molecules, effective spin-spin exchange interactions are obtained. Due to the large transition dipole moments of Rydberg atoms and negligible molecular permanent dipole moments, the resulting central spin model features strong central-bath $XX$ interactions, and no bath-bath interactions.

The system can be realized, \emph{e.g.}, by using optical tweezers to precisely position the particles, which allows us to control the coupling parameter for each bath spin separately. In particular, arranging the molecules in a ring shape allows us to smoothly tune the system between fully homogeneous and increasingly inhomogeneous interactions, simply by changing the angle of an external field that defines the quantization axis. To illustrate this capability, we show example scenarios where dynamics depend directly on this angle parameter. In the first scenario, we consider a simulated decoherence process of a qubit interacting with an environment, where the angle parameter controls the decoherence timescale. In the second scenario, we demonstrate a simple communication protocol, in which a classical bit is sent between two bath spins that share the same strength of interaction with the central spin. Here the angle parameter controls the resulting transfer time.

The setup should be feasible to realize in present ultracold physics laboratories. Although for the ring-shaped tweezer layout the maximum possible molecule number is limited, other types of particle geometries, \emph{e.g.}, spherical three-dimensional layouts, could allow for much larger molecule numbers. The setup may therefore be useful for simulating systems with number of bath spins high enough to be impractical for exact numerical simulation.

The proposed setup admits several potential directions for further exploration and improvement. For example, using Rydberg atoms as bath spins in place of molecules would change the Hamiltonian significantly, possibly allowing to reach different parameter regimes, \emph{e.g.} with non-negligible bath-bath interactions. Other interesting possibilities include systems where more than one atom is trapped in a single optical tweezer (see, \emph{e.g.}, Refs.~\cite{2019-Liu-PRX,2020-Hood-PRR,2022-Brooks-NJP}), corresponding to models with more than one central spin.

Using alternative molecule or atom species might also help extend the capabilities of the simulator.
For example, molecules with particularly strong internuclear dipole moments (such as certain dimers containing silver or copper atoms~\cite{2021-Smialkowski-PRA}) might allow us to realize non-negligible bath-bath interactions, especially if the intermolecular distances are made as small as possible.
Similarly, using molecules with large dipole moments $d$ and small rotational constants $B_\mathrm{rot}$ (which maximizes the permanent dipole moment induced by an electric field) might allow realizing nontrivial effective $\hat{S}^{(0)}_z \hat{S}^{(k)}_z$ interactions, opening the way to simulating $XXZ$ models beyond the $XX$ model.

A technique not explored here is using external microwave ac radiation to dress molecular states, which allows us to obtain dressed pseudospin states of desired composition. Such an approach is taken in many works describing simulated spin systems realized with ultracold molecules~\cite{2006-Micheli-NatPhys,2011-Gorshkov-PRA,2011-Gorshkov-PRL,2013-Gorshkov-MolPhys,2013-Manmana-PRB,2015-Wall-NJP}, and may be applicable to this hybrid atom--molecule system as well. However, the microwave radiation would affect both the molecules and the atom, leading to potential complications. Nevertheless, if successfully implemented, this technique might enable even greater control over the system.

\begin{acknowledgements}
Financial support from the Foundation for Polish Science within the First Team programme cofinanced by the European Union under the European Regional Development Fund is gratefully acknowledged. The computational part of this research has been partially supported by the PL-Grid Infrastructure.
\end{acknowledgements}

\appendix

\section{Atomic and molecular levels}
\label{sec:Spectra}

In this section, we describe in more detail the methods used to obtain the single-particle eigenspectra of the molecular Hamiltonians $\hat{h}_\mathrm{mol}^{(k)}$ and the Rydberg atom Hamiltonian $\hat{h}_\mathrm{Ryd}$. Additionally, we discuss the extent to which external fields can be employed to adjust the individual level energies and interlevel transition frequencies.

\begin{figure*}[t]
 \centering
 \includegraphics[width=1.00\textwidth]{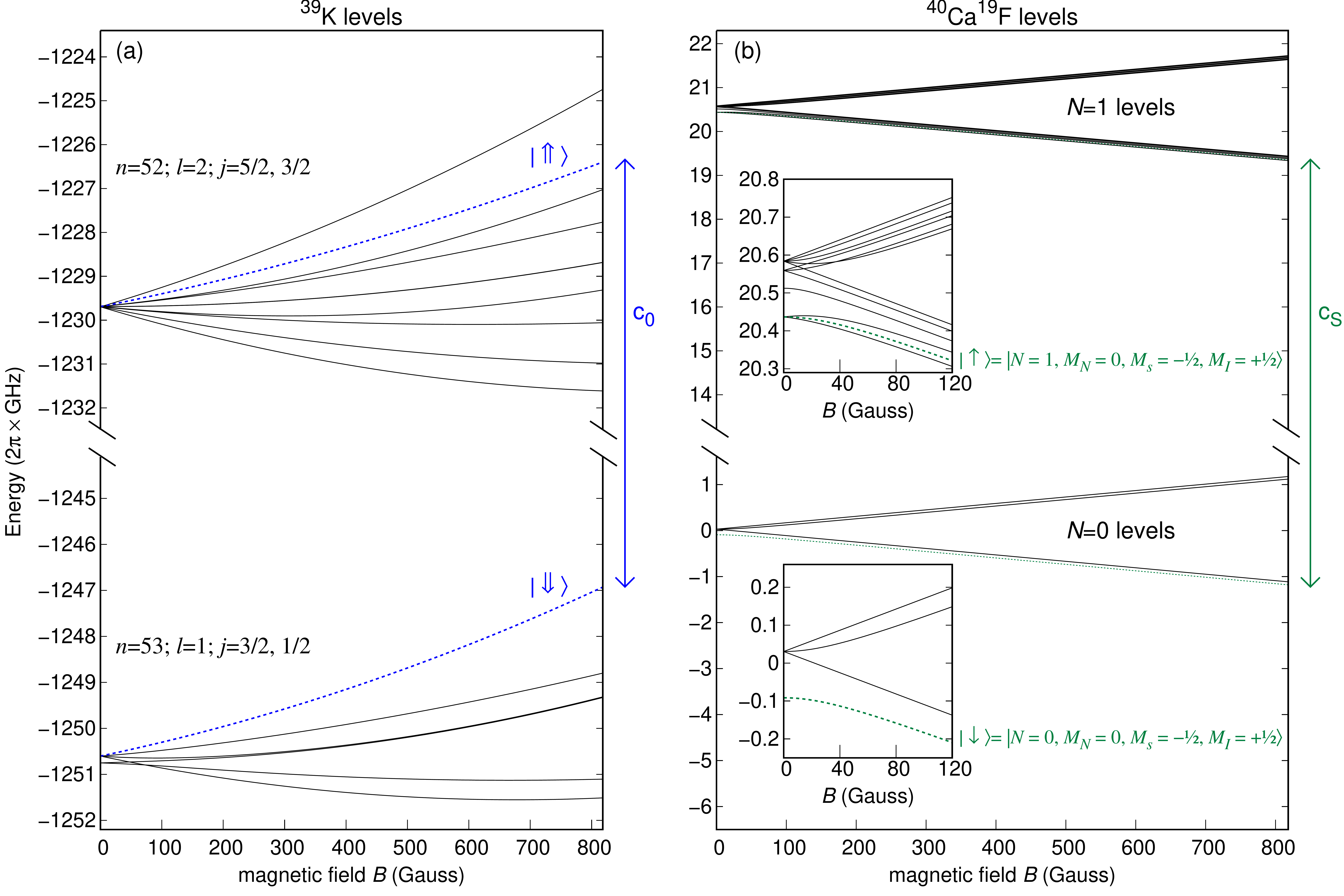}
 \caption{The atomic and molecular energy spectrum as a function of magnetic field $B$. We use the same atomic species ($^{39}$K) and molecular species ($^{40}$Ca$^{19}$F) as the example setup described in Section~\ref{sec:Model-ExampleChoice}. (a) The $n=52, l=2$ levels and the $n=53, l=1$ levels of $^{39}$K. The blue dashed lines indicate the two levels which, in Section~\ref{sec:Model-ExampleChoice}, are chosen to represent central spin states $|\Uparrow\rangle$ and $|\Downarrow\rangle$. The energy difference between them, at a particular magnetic field $B = 818.1\,\mathrm{Gauss}$, is indicated by $c_0$. Energies are calculated with the \textsc{pairinteraction} Python package, and are given relative to the ionization threshold. (b) The hyperfine-split $N=0$ and $N=1$ levels of the $^{40}$Ca$^{19}$F molecule. Insets show in greater detail the spectrum at smaller magnetic fields. The green dashed lines indicate the two levels which, in Section~\ref{sec:Model-ExampleChoice}, are chosen to represent bath spin states $|\uparrow\rangle$ and $|\downarrow\rangle$. We also show the values of the $N,M_N,M_s,M_I$ quantum numbers which characterize these pseudospin levels at higher magnetic fields. The energy difference between these levels, at a particular magnetic field $B = 818.1\,\mathrm{Gauss}$, is indicated by $c_S$. Energies are calculated by direct numerical diagonalization of Eq.~\eqref{eq:total_molecular_hamiltonian} using the parameters from Table~\ref{tab:CaF_hyperfine_parameters}. The magnetic field is chosen so that $c_0$ and $c_S$ are almost equal, allowing for resonant energy exchange between atoms and molecules.}
 \label{fig:K-CaF-Zeeman-levels}
\end{figure*}

As a concrete example, in Fig.~\ref{fig:K-CaF-Zeeman-levels}, we depict in detail the energy spectrum of atom $^{39}$K [Fig.~\ref{fig:K-CaF-Zeeman-levels}(a)] and molecule $^{40}$Ca$^{19}$F [Fig.~\ref{fig:K-CaF-Zeeman-levels}(b)] as a function of the magnetic field. The figure illustrates the example setup described in Section~\ref{sec:Model-Species}, in which specific levels of $^{39}$K and $^{40}$Ca$^{19}$F, at a high magnetic field, are chosen to represent the central spin states $\Uparrow,\Downarrow$ and the bath spin states $\uparrow,\downarrow$ respectively. In Fig.~\ref{fig:K-CaF-Zeeman-levels}, we indicate these pseudospin states with thicker lines.

\subsection{The CaF hyperfine Hamiltonian}
\label{sec:CaFSpectrum}

In this section, we examine in detail the internal molecular Hamiltonian $\hat{h}_\mathrm{mol}^{(k)}$, showing how it can be numerically diagonalized to find the molecular levels for any strength of external fields.

We focus on an example molecule species $^{40}$Ca$^{19}$F, a polar rigid-rotor molecule, in its ground vibrational and electronic ($^2\Sigma$) state. In the $^2 \Sigma$ state, the molecule angular momentum comes from three sources: the rotational angular momentum ${ N=0,1,2,\ldots }$, the spin of the unpaired electron $s=1/2$, and the spin of the $^{19}$F nucleus $I=1/2$ (the $^{40}$Ca nucleus has zero spin). Using these angular momenta, we can define an uncoupled basis of states ${ |N,M_N,s,M_s,I,M_I\rangle }$, where $M_N,M_s,M_I$ are the projections of the corresponding angular momenta on the space-fixed quantization axis. This is a convenient basis for diagonalizing the molecular Hamiltonian. Note that changing the values of $s$ and $I$ requires energy scales far beyond those considered in our model, therefore we treat them as fixed values, and write the uncoupled basis states more concisely as ${ |N,M_N,M_s,M_I\rangle }$.

The internal Hamiltonian for a single molecule can be written as
\begin{equation}
\label{eq:total_molecular_hamiltonian}
\hat{h}_\mathrm{mol} = \hat{h}_\mathrm{rot} + \hat{h}_\mathrm{hf} + \hat{h}_\mathrm{E;mol} + \hat{h}_\mathrm{B;mol},
\end{equation}
where the rotational and hyperfine terms $\hat{h}_\mathrm{rot} + \hat{h}_\mathrm{hf}$ describe the zero-field levels, while $\hat{h}_\mathrm{E;mol} + \hat{h}_\mathrm{B;mol}$ describe external field interactions further dressing the eigenstates. Each term is described below in more detail. The values of all the physical parameters mentioned below are listed in Table~\ref{tab:CaF_hyperfine_parameters}.

The first term, $\hat{h}_\mathrm{rot}$, describes the rotational energy. For a rigid-rotor molecule, it is given by
\begin{equation}
 \hat{h}_\mathrm{rot} = B_\mathrm{rot} \hat{N}^2,
\end{equation}
with $B_\mathrm{rot}$ being the rotational constant of the molecule.

The second term, $\hat{h}_\mathrm{hf}$, describes the hyperfine couplings between angular momenta. For $^{40}$Ca$^{19}$F it takes the form
\begin{align}
 \hat{h}_\mathrm{hf} = &\gamma \vec{s} \cdot \vec{N} + (b + c/3) \vec{s} \cdot \vec{I} \nonumber \\
 &+ (c/3) \sqrt{6} T^2(\vec{C}) \cdot T^2(\vec{I},\vec{s}) + c_F \vec{I} \cdot \vec{N},
\end{align}
where the terms are, in order: the electron-spin--rotation interaction; the scalar and tensor parts of the nuclear-spin--electron-spin interaction; and the nuclear-spin--rotation interaction. The magnitude of the hyperfine splittings is controlled by the coupling constants $\gamma, b, c$, and $c_F$. The resulting hyperfine splittings are on the order of $ \sim 2\pi \times 10^7\,\mathrm{Hz}$, which is much larger than the dipolar interaction energies in our system, and so the hyperfine couplings cannot be neglected. A more detailed overview of the hyperfine couplings in $^2 \Sigma$ molecules can be found in many sources, \emph{e.g.} in Refs.~\cite{2003-Brown-Book,2018-Aldegunde-PRA}.

The terms $\hat{h}_\mathrm{E;mol}$ and $\hat{h}_\mathrm{B;mol}$ describe the interaction with external electric and magnetic fields, respectively. For fields directed along the space-fixed quantization axis $\vec{e}_0$, these terms take simple forms:
\begin{align}
  \hat{h}_\mathrm{E;mol} &= -E_\mathrm{dc} \hat{d}_0, \\
  \hat{h}_\mathrm{B;mol} &= B (\mu_B g_s \hat{M}_s - \mu_N g_I \hat{M}_I - \mu_B g_r \hat{M}_N),
\end{align}
where $g_s, g_I, g_r$ are the g-factors corresponding to the electron spin, $^{19}$F nuclear spin, and rotation, respectively. $\mu_B$ and $\mu_N$ are the standard Bohr magneton and nuclear magneton. We neglect the small screening factor~\cite{2003-Brown-Book} that normally modifies the magnetic field.

\begin{table}[t]
\begin{tabular}{ll}
\hline
\hline
$I$ & $1/2$~\cite{2003-Brown-Book} \Tstrut\\
$s$ & $1/2$ \\
$B_\mathrm{rot}$ & $2\pi\times10267.539\,\mathrm{MHz}$~\cite{1994-Anderson-APJ}\\
$d$ & $3.07\,\mathrm{debye}$~\cite{1984-Childs-JChemPhys} \Bstrut\\
\hline
$\gamma$ & $2\pi\times39.498\,\mathrm{MHz}$~\cite{1981-Childs-JMolSpec} \Tstrut\\
$b$ & $2\pi\times108.476\,\mathrm{MHz}$~\cite{1981-Childs-JMolSpec}\\
$c$ & $2\pi\times40.647\,\mathrm{MHz}$~\cite{1981-Childs-JMolSpec}\\
$c_F$ & $2\pi\times0.029\,\mathrm{MHz}$~\cite{1981-Childs-JMolSpec} \Bstrut\\
\hline
$g_s$ & $2.0023$ \Tstrut\\
$g_I$ & $5.2545$~\cite{1991-CRC-Book} \\
$g_r$ & $-5.13 \times 10^{-5}$~\cite{2020-Caldwell-PRL-Long} \Bstrut\\
\hline
\hline
\end{tabular}
\caption{Physical parameters of the $^{40}$Ca$^{19}$F molecular Hamiltonian used for numerical calculations in this paper.}
\label{tab:CaF_hyperfine_parameters}
\end{table}

To numerically find the energy levels at any $B$ or $E_\mathrm{dc}$, first we express the Hamiltonian in Eq.~\eqref{eq:total_molecular_hamiltonian} as a matrix in the ${ |N,M_N,M_s,M_I\rangle }$ basis. Including all the possible values of $M_N,M_s,M_I$, there are ${ (2N+1) \times 2 \times 2 }$ such basis states for each $N$. In our calculation the basis includes all states up to $N=2$. This is necessary to obtain accurate results for $N=0,N=1$ molecular levels at nonzero electric fields, which cause couplings to higher $N$.

We now briefly recall the relations necessary to calculate the matrix elements for all the molecular Hamiltonian terms. (For brevity, in Eqs.~\eqref{eq:appendix-matrix-elements-1}--~\eqref{eq:appendix-product-tensor} any conserved quantum numbers are left out of the kets.)
\begin{align}
 \hat{N}^2|N,M_N\rangle &= N(N+1)|N,M_N\rangle, \label{eq:appendix-matrix-elements-1}\\
 \hat{M}_N|N,M_N\rangle &= M_N|N,M_N\rangle, \label{eq:appendix-matrix-elements-2}\\
 \hat{N}_\pm|N,M_N\rangle &= \sqrt{N(N+1) - M_N(M_N \pm 1)}|N,M_N\pm1\rangle, \label{eq:appendix-matrix-elements-3}\\
 \vec{s} \cdot \vec{N} &= M_s M_N + \frac{1}{2}(s_+ N_- + s_- N_+). \label{eq:appendix-matrix-elements-4}
 \end{align}
 Analogous relationships as Eqs.~\eqref{eq:appendix-matrix-elements-1}--~\eqref{eq:appendix-matrix-elements-4} hold for other angular momenta and their dot products.

One term which requires separate attention is the tensor part of the nuclear-spin--electron-spin interaction. It involves a scalar product between two rank-two tensors, which correspond respectively to spherical harmonics and to the product of angular momenta $\vec{I},\vec{s}$. The corresponding matrix elements can be found by the following equations:
\begin{align}
 T^2(\vec{C}) \cdot T^2(\vec{I},\vec{s}) &= \sum_{p=-2}^{p=+2} (-1)^p T^2_p(\vec{C}) T^2_{-p}(\vec{I},\vec{s}), \\
 T^2_p(\vec{C}) &= C^2_p(\theta,\phi),
 \end{align}
 \begin{align}
  &\langle N,M_N|C^j_p(\theta,\phi)|N',M'_N\rangle = (-1)^{M_N} \sqrt{(2N+1)(2N'+1)} \nonumber\\
 &\times \begin{pmatrix}
  N & j & N' \\
  -M_N & p & M'_N
\end{pmatrix} \begin{pmatrix}
  N & j & N' \\
  0 & 0 & 0
\end{pmatrix},
\label{eq:appendix-spherical-harmonic}
\end{align}
\begin{align}
&\langle M_s,M_I|T^2_p(\vec{I},\vec{s})|M'_s,M'_I\rangle = (-1)^{I-M_I+s-M_s-p} \nonumber \\
&\times \sqrt{5 I (I+1) (2I+1) s (s+1) (2s+1)} \nonumber\\
&\times \sum_{p'=-1}^{+1} \left[ \begin{pmatrix}
  1 & 1 & 2 \\
  p' & p-p' & -p
\end{pmatrix} \begin{pmatrix}
  I & 1 & I \\
  -M_I & p' & M'_I
\end{pmatrix} \right. \nonumber\\
&\times \left. \begin{pmatrix}
  s & 1 & s \\
  -M_s & p-p' & M'_s
\end{pmatrix} \right],
\label{eq:appendix-product-tensor}
\end{align}
where the expressions in round brackets are the usual Wigner 3-j symbols. The above expressions for tensor elements can be found in a number of sources; a convenient summary is available, \emph{e.g.}, in Appendix A of Ref.~\cite{2011-Gorshkov-PRA}.

The term $\hat{h}_\mathrm{E;mol}$ contains the electric dipole operator $\hat{d}_q$, whose matrix elements can be found by relating it to the rank-1 spherical harmonics tensor,
\begin{equation}
\label{eq:appendix-dipole-operator}
\hat{d}_q = d C^1_q(\theta,\phi).
\end{equation}
Here $d$ is the molecule's internuclear dipole moment. Based on Eqs.~\eqref{eq:appendix-spherical-harmonic},~\eqref{eq:appendix-dipole-operator}, the $\hat{d}_q$ operator selection rules can be expressed as $\Delta N=\pm1, \Delta M_N = q, \Delta M_s = 0, \Delta M_I = 0$.

With all the matrix elements known, the Hamiltonian can be diagonalized using standard numerical methods, \emph{e.g.} the \textsc{LAPACK} diagonalization routines. The resulting energy levels, as a function of the magnetic field, are depicted in Fig.~\ref{fig:K-CaF-Zeeman-levels}(b).

In the low-field limit, the molecular levels are linear combinations of different ${ |N,M_N,M_s,M_I\rangle }$ states, and $M_N,M_s,M_I$ are not good quantum numbers. Instead, the low-field states are best described in the coupled representation ${ |N,F,M_F\rangle }$; in this limit the good quantum numbers are $N,s,I$, the total angular momentum (sum of the three angular momenta) $F = 0,1,2,\ldots$, and its quantization-axis projection $M_F = -F, -F+1, \ldots, +F$. The low-field levels are split into groups with different $N$, separated by energies $\sim B_\mathrm{rot}$. Within each $N$ group, the levels are further grouped into manifolds corresponding to different $F$, separated by energies on the order of hyperfine coupling constants ($\sim 2\pi \times 10\,\mathrm{MHz}$ in case of $^{40}$Ca$^{19}$F). In Fig.~\ref{fig:K-CaF-Zeeman-levels}(b), it can be seen the low-field $N=0$ levels are grouped into two manifolds (they correspond, in order of energy, to $F=0$ and $F=1$), while the low-field $N=1$ levels are grouped into four manifolds (they correspond to $F=1$, $F=0$, $F=1$, $F=2$; note there are two different manifolds with $F=1$). Each manifold contains $2F+1$ levels with different $M_F$, which in the zero-field limit are degenerate.

At higher magnetic fields, the levels undergo increasing Zeeman shifts and the degeneracy of energies within each $F$ manifold is lifted. Because of the unpaired electron spin, which couples strongly to $B$, the Zeeman shifts in CaF can be quite substantial in comparison to hyperfine shifts; for the depicted magnetic-field strengths, the level energies change by values of $\sim 2\pi\times1\,\mathrm{MHz}$ per $1\,\mathrm{G}$. At a high enough field, where the Zeeman term dominates over the hyperfine couplings ($\approx 100\,\mathrm{G}$ in the case of $^{40}$Ca$^{19}$F), $F$ ceases to be a good quantum number. All the high-field levels instead have well-defined $M_N,M_s,M_I$, and are very well approximated by the uncoupled basis states ${ |N,M_N,M_s,M_I\rangle }$ ($M_F = M_N+M_s+M_I$ also remains a good quantum number).

The Zeeman shift can be used to adjust the frequency of an electric dipole transition between two molecular states. However, the magnitude of the transition frequency shift is typically much smaller than the Zeeman shifts of the individual state energies. For example, in $^{40}$Ca$^{19}$F at low magnetic fields, the transition frequencies between two $N=0$, $N=1$ levels are shifted from the zero-field value by less than $\sim 2\pi\times1\,\mathrm{MHz}$ per $1\,\mathrm{G}$. This is very small compared with the transition frequency $2 B_\mathrm{rot} \sim 2\pi\times 20\,\mathrm{GHz}$. For higher values of $B$, where the individual levels are given by ${ |N,M_N,M_s,M_I\rangle }$, the transition frequencies saturate and become almost insensitive to further increase of $B$. This is because the electric dipole operator $\hat{d}_q$ only couples states which have the same values of $M_s$ and $M_I$, and which therefore exhibit almost identical linear slopes of Zeeman shifts (except for the very small part proportional to $M_N$). As an example, in $^{40}$Ca$^{19}$F, the frequencies of $N=0 \leftrightarrow N=1$ transitions are shifted by only $\sim 10\,\mathrm{MHz}$ at a magnetic field $\approx 100\,\mathrm{G}$, and they remain almost unchanged with a further increase of $B$.

Electric fields also can shift molecular energies. The Stark shift of a molecular level in a rigid-rotor molecule is on the order of $\sim (d E_\mathrm{dc})^2/(2 B_\mathrm{rot})$, and depends significantly on $N$ and $M_N$~\cite{2003-Brown-Book}. However, for an electric field up to 50 V/cm (the approximate field-ionization limit of the Rydberg atom at $n=50$), these Stark shifts are typically very small in comparison to the rotational transition frequency. For example, at $E_\mathrm{dc} = 50\,\mathrm{V/cm}$, $^{40}$Ca$^{19}$F energies are shifted by a few hundred kHz at most.

\subsection{The K Hamiltonian spectrum}
\label{sec:KSpectrum}

In this section, we examine the spectrum of the internal atomic Hamiltonian $\hat{h}_\mathrm{Ryd}$. Unlike the previous section on $\hat{h}_\mathrm{mol}$, we will not delve into the process of diagonalizing $\hat{h}_\mathrm{Ryd}$ in detail, since we rely on pre-existing libraries such as \textsc{ARC} and \textsc{pairinteraction} to obtain the atomic levels and their energies at various electric or magnetic fields. Instead, we will briefly discuss the efficiency of adjusting atomic energies with external fields.

As an example, Fig.~\ref{fig:K-CaF-Zeeman-levels}(a) illustrates the atomic level spectrum of $^{39}$K as a function of the magnetic field.  The energies depicted in the figure were calculated using the \textsc{pairinteraction} library. Comparing Figs.~\ref{fig:K-CaF-Zeeman-levels}(a) and ~\ref{fig:K-CaF-Zeeman-levels}(b), it can be seen the atomic Zeeman shifts at approx. $B = 800\,\mathrm{G}$ can be noticeably larger than molecular shifts. This is due to the diamagnetic atomic Zeeman term, which is quadratic in $B$ and becomes significant at higher fields. We also note the atomic Stark shifts are similarly much stronger than molecular Stark shifts, because of the much larger electric polarizability of a Rydberg atom, scaling as $\sim n^7$~\cite{1994-Gallagher-Book}.

It is also worth noting that the Zeeman shift can change electric dipole transition frequencies in atoms much more strongly than in molecules. This is because such transitions occur between levels with different orbital angular momenta $l$, and the orbital angular momentum significantly affects the magnitude of the atomic Zeeman shift~\cite{2017-Weber-JPB}. For example, at $B \approx 1000\,\mathrm{G}$, the transition frequency between two $n=53, l=1$ and $n=52, l=2$ states can be changed from the zero-field value by values on the order of several hundred $\sim 2\pi \times \mathrm{MHz}$.

\section{Modifying the phases of interaction coefficients}
\label{sec:Phase-Transformation}

In the spin Hamiltonian $\hat{H}_\mathrm{eff}$ [Eq.~\eqref{eq:eff-spin-hamiltonian}], the interaction coefficients $C_k$ [Eq.~\eqref{eq:effective-interaction-coefficient}] can be complex numbers. However, it is often desirable to have control over their phases. For example, descriptions of qubit decoherence in quantum dots typically assume real-valued interactions. In this section, we describe how to effectively modify the phases of $C_k$ in the context of time evolution.

Suppose we want to evolve an initial state ${ |\Psi_0\rangle }$ under a Hamiltonian having the same form as $\hat{H}_\mathrm{eff}$, but with interaction parameters modified as $C_k \rightarrow C_k e^{-i \xi_k}$ (with some desired values of $\xi_k$). To do so, first define the following unitary transformation
\begin{equation}
\label{eq:unitary-transformation-f}
\hat{\mathcal{F}} = \hat{1}^{(0)} \otimes \hat{\mathcal{F}}^{(1)} \otimes \hat{\mathcal{F}}^{(2)} \otimes \ldots \otimes \hat{\mathcal{F}}^{(N_\mathrm{bath})},\\
\end{equation}
where
\begin{equation}
\label{eq:unitary-transformation-f-factors}
\hat{\mathcal{F}}^{(k)} = {e^{i \xi_k} |\uparrow^{(k)}\rangle\langle \uparrow^{(k)} | + |\downarrow^{(k)}\rangle\langle\downarrow^{(k)}|}.
\end{equation}
This transformation can be applied to the Hamiltonian to modify the phases of $C_k$:
\begin{align}
\label{eq:modified-phases-hamiltonian}
 \hat{\mathcal{F}} \hat{H}_\mathrm{eff} \hat{\mathcal{F}}^\dagger &= c_0 \hat{S}^{(0)}_z + c_S \sum_{k=1}^{N_\mathrm{bath}} \hat{S}^{(k)}_z \nonumber \\
 &+ \sum_{k=1}^{N_\mathrm{bath}} ( C_k e^{-i \xi_k} \hat{S}^{(0)}_+ \hat{S}^{(k)}_- + C^*_k e^{i \xi_k} \hat{S}^{(0)}_- \hat{S}^{(k)}_+ ).
\end{align}

Since $\hat{\mathcal{F}}$ is unitary, $e^{-i \hat{\mathcal{F}} \hat{H} \hat{\mathcal{F}}^\dagger t} = \hat{\mathcal{F}} e^{-i \hat{H} t} \hat{\mathcal{F}}^\dagger$.
Therefore, for the time evolution of any initial state ${ |\Psi_0\rangle }$, the following holds:
\begin{equation}
 e^{-i \hat{\mathcal{F}} \hat{H}_\mathrm{eff} \hat{\mathcal{F}}^\dagger t} |\Psi_0\rangle = \hat{\mathcal{F}} e^{-i \hat{H}_\mathrm{eff} t} (\hat{\mathcal{F}}^\dagger |\Psi_0\rangle).
\end{equation}

Therefore, the evolution of the initial state ${ |\Psi_0\rangle }$ under the modified Hamiltonian [Eq.~\eqref{eq:modified-phases-hamiltonian}] can be obtained by setting a modified initial state ${ \hat{\mathcal{F}}^\dagger |\Psi_0\rangle }$, letting it evolve under the original Hamiltonian $\hat{H}_\mathrm{eff}$, and acting with $\hat{\mathcal{F}}$ on the resulting evolved state ${ |\Psi(t)\rangle }$. On the experimental level, this means simply adjusting the initial-state preparation and adding an extra step to the result analysis. Thus, the phases of all $C_k$ can be regarded as freely controllable parameters. In particular all the $C_k$ can be changed to be real and positive by setting $\xi_k = \arg C_k$.

It is worth noting that, under certain initial states, the time evolution of some observables is insensitive to phases of $C_k$ in the first place.
Consider the time evolution of an observable described by some operator $\hat{O}$. Under the Hamiltonian $\hat{H}_\mathrm{eff}$, the value of $O$ evolves as
\begin{equation}
\label{eq:O-evolution-normal}
 O(t) = \langle \Psi_0 | e^{i \hat{H}_\mathrm{eff} t} \hat{O} e^{-i \hat{H}_\mathrm{eff} t} | \Psi_0 \rangle.
\end{equation}

Now consider its evolution under the Hamiltonian with modified interaction coefficients [Eq.~\eqref{eq:modified-phases-hamiltonian}]:
\begin{align}
 O(t) &= \langle \Psi_0 | e^{i \hat{\mathcal{F}} \hat{H}_\mathrm{eff} \hat{\mathcal{F}}^\dagger t} \hat{O} e^{-i \hat{\mathcal{F}} \hat{H}_\mathrm{eff} \hat{\mathcal{F}}^\dagger t} | \Psi_0 \rangle \nonumber \\
 &= \langle \Psi_0 | \hat{\mathcal{F}} e^{i  \hat{H}_\mathrm{eff}  t} \hat{\mathcal{F}}^\dagger \hat{O}  \hat{\mathcal{F}} e^{-i  \hat{H}_\mathrm{eff}  t} \hat{\mathcal{F}}^\dagger | \Psi_0 \rangle. \label{eq:O-evolution-modified}
\end{align}

The right-hand sides of Eqs.~\eqref{eq:O-evolution-normal} and ~\eqref{eq:O-evolution-modified} are the same (\emph{i.e.}, the evolution of $O(t)$ remains identical regardles of the phases of $C_k$) as long as two conditions are fulfilled. First, the initial state ${ |\Psi_0\rangle }$ must be an eigenstate of $\hat{\mathcal{F}}$, so that the operation ${ \hat{\mathcal{F}}^\dagger |\Psi_0\rangle }$ simply multiplies it by a phase factor. This is the case, \emph{e.g.}, for a single product state, such as ${ |\Psi_0\rangle } = { |\Uparrow;\downarrow^{(1)}, \uparrow^{(2)}, \ldots ,\downarrow^{(N_\mathrm{bath})} \rangle }$. Second, the operator $\hat{O}$ must commute with $\hat{\mathcal{F}}$, so that $\hat{\mathcal{F}}^\dagger \hat{O} \hat{\mathcal{F}} = \hat{O}$. This is the case, \emph{e.g.}, for operators $\hat{S}^{(0)}_z$ or $\hat{S}^{(k)}_z$.

A similar reasoning can also apply to a randomly chosen initial state which has the form
\begin{equation}
\label{eq:random-state-appendix}
 |\Psi_0\rangle = \sum_j \alpha_j |j\rangle,
\end{equation}
where the sum runs over many basis states ${ |j\rangle }$ (all the possible basis states, or limited to some subset, \emph{e.g.}, with given total spin), and all the coefficients $\alpha_j$ are complex numbers, having independently random phases picked from an uniform distribution.
	The operation ${ \hat{\mathcal{F}}^\dagger | \Psi_0 \rangle }$ modifies the phase of each $\alpha_j$ in a way that is indistinguishable from picking another random choice for each $\arg{\alpha_j}$.
	Therefore, using a Hamiltonian with different phases of $C_k$ results in the same outcome as replacing the random initial state with another, equally random one.

\section{Alternative way of defining the pseudospin basis}

\subsection{Using a linear combination of molecular states as a pseudospin state}

\label{sec:LinearCombination}

In the main text (Section~\ref{sec:Model-EffectiveSpinH}), we have shown how two specific molecular levels can be used as pseudospin basis states. There is also an alternative possibility of defining the pseudospin basis, which appears if the molecular spectrum has two or more degenerate levels (as is the case in the zero-field limit). In this case, it is possible to define an effective pseudospin state as a linear combination of these degenerate levels.

As an example, suppose that we have defined a pair of atomic levels as ${ |\Downarrow\rangle,|\Uparrow\rangle }$, and defined ${ |\downarrow\rangle }$ as some molecular level ${ |N=0,F,M_F\rangle }$ which is not degenerate with any other levels. Suppose also there exist several degenerate $N=1$ molecular levels ${ |u_1\rangle,|u_2\rangle,|u_3\rangle,\ldots }$. Finally, assume that $\hat{V}_\mathrm{atom-mol}^{(k)}$ has nonzero matrix elements between the product states ${ |\Uparrow ; \downarrow^{(k)}\rangle \leftrightarrow |\Downarrow ; u_i^{(k)}\rangle }$, while couplings to other states are zero or off-resonant. Under these conditions, the action of $\hat{V}_\mathrm{atom-mol}^{(k)}$ on ${ | \Uparrow; \downarrow^{(k)} \rangle }$ is given by
\begin{align}
&\hat{V}_\mathrm{atom-mol}^{(k)} | \Uparrow; \downarrow^{(k)} \rangle \nonumber \\
&= \sum_{u_i} |\Downarrow ; u_i^{(k)}\rangle \left[ \langle \Downarrow ; u_i^{(k)} | \hat{V}_\mathrm{atom-mol}^{(k)} | \Uparrow ; \downarrow^{(k)} \rangle \right] \nonumber \\
&\equiv C^*_k |\Downarrow ; \uparrow^{(k)}\rangle,
\end{align}
so that the up-spin state is given by
\begin{equation}
\label{eq:linear_combination_up_spin_state}
|\uparrow^{(k)}\rangle = \sum_{u_i} \frac{\langle \Downarrow ; u_i^{(k)} | \hat{V}_\mathrm{atom-mol}^{(k)} | \Uparrow; \downarrow^{(k)} \rangle}{C^*_k} |u_i^{(k)}\rangle.
\end{equation}
The magnitude of the interaction parameter $C^*_k$ can then be found by normalizing ${ |\uparrow^{(k)}\rangle }$ to unity. It is convenient to simply take $C^*_k$ as a real and positive normalization factor, so that
\begin{equation}
\label{eq:linear_combination_interaction_coefficient}
C^*_k = C_k = \sqrt {\sum_{u_i} |\langle \Downarrow ; u_i^{(k)} | \hat{V}_\mathrm{atom-mol}^{(k)} | \Uparrow ;\downarrow^{(k)} \rangle|^2}.
\end{equation}

Therefore we can define an effective pseudospin state ${ |\uparrow^{(k)}\rangle }$ as a linear combination of ${ |u_i^{(k)}\rangle }$, with a composition different for each molecule $k$ (since the dipole interaction matrix elements depend on the molecule positions). It is easy to check that $\hat{V}_\mathrm{atom-mol}^{(k)}$ couples the states ${ | \Uparrow; \downarrow^{(k)} \rangle }$ and ${ | \Downarrow; \uparrow^{(k)} \rangle }$ only to each other. Therefore, each molecule $k$ remains in the subspace of states ${ |\downarrow^{(k)}\rangle,|\uparrow^{(k)}\rangle }$ throughout the time evolution, mimicking a two-level system. Note that in this approach, $C_k$ is defined as always real and positive, and complex values in the spin interaction term are instead hidden in the coefficients ${ \langle u^{(k)}_i |\uparrow^{(k)}\rangle }$.

In the main text, our derivation of the effective Hamiltonian assumes that ${ |\uparrow^{(k)}\rangle,|\downarrow^{(k)}\rangle }$ are both chosen as eigenstates of $\hat{h}_\mathrm{mol}^{(k)}$. If all the levels ${ |u^{(k)}_i\rangle }$ have exactly identical energies, their linear combination is indeed an eigenstate of $\hat{h}_\mathrm{mol}^{(k)}$. Even if the energies of ${ |u^{(k)}_i\rangle }$ are not identical, the scheme should still work, as long as the energy differences are sufficiently small. Assume the maximum considered duration of the system evolution is $\tau_\mathrm{max}$. As long as the energy differences are $\ll 1/\tau_\mathrm{max}$, the individual components of ${ |\uparrow^{(k)}\rangle }$ remain in phase within the time $\tau_\mathrm{max}$, and ${ |\uparrow^{(k)} \rangle }$ approximately behaves like an eigenstate. We note this scheme is conveniently combined with using electric fields to tune the mismatch $c_\Delta$. A small electric field modifies mostly the atomic energies, leaving the molecule energies almost unchanged, and nearly completely preserving the degeneracy of ${ |u^{(k)}_i\rangle }$.

Note that this scheme only works because the molecule-molecule interactions are negligible. This is because the composition of ${ |\uparrow^{(k)}\rangle }$ is defined by the angles $\theta_k,\phi_k$ relative to the atom, while molecule-molecule interactions would act at many different angles, and so would not reliably preserve the composition of ${ |\uparrow^{(k)}\rangle }$.

An advantage of this approach is that it allows us to realize $C_k$ with forms otherwise not available. In particular (as we show for a specific example in Appendix~\ref{sec:LinearCombinationCaFExample}), it allows to realize a system where $C_k$ does not cross zero regardless of the particle positions, which may be desirable in some cases. On the other hand, this approach complicates the preparation of a desired initial state. It is difficult to prepare a molecule in a specific linear combination of degenerate levels, with precisely defined amplitudes and phases. However, even with this complication, certain initial states might be attainable, \emph{e.g.}, by setting up the initial positions, exciting the atom into the state ${ |\Uparrow\rangle }$, letting it transfer its excitation to the molecules, and measuring the atom state at a specific time to fix the molecule states~\cite{2008-Ferraro-EPJSpecTop}.

\subsection{Example of realization for CaF system}

\label{sec:LinearCombinationCaFExample}

To demonstrate the above approach, here we consider an example system, composed of $^{40}$Ca$^{19}$F molecules as bath spins, and an $^{85}$Rb atom as the central spin. We use an analogous presentation as in Section~\ref{sec:Model-ExampleChoice} of the main text. The electric field is set to $E_\mathrm{dc}^\mathrm{res} = 3042.5\,\mathrm{mV/cm}$, while the magnetic field is zero. The molecular down-spin state is chosen as
\begin{equation}
 |\downarrow^{(k)}\rangle = |(\overline{N=0,F=0,M_F=0})^{(k)}\rangle.
\end{equation}
The molecular up-spin state ${ |\uparrow^{(k)}\rangle }$ is defined as a linear combination of three $N=1$ states, which we denote as ${ |u_0^{(k)}\rangle },{ |u_{+1}^{(k)}\rangle },{ |u_{-1}^{(k)}\rangle }$:
\begin{align}
 |u_0^{(k)}\rangle &\equiv |(\overline{N=1,F=1^-,M_F=0})^{(k)}\rangle \\
 |u_{+1}^{(k)}\rangle &\equiv |(\overline{N=1,F=1^-,M_F=+1})^{(k)}\rangle \\
 |u_{-1}^{(k)}\rangle &\equiv |(\overline{N=1,F=1^-,M_F=-1})^{(k)}\rangle.
\end{align}
In zero field, these states have exactly identical energies. In the small electric field $E_\mathrm{dc}^\mathrm{res}$, the energy of the $M_F=0$ state differs by approximately $2\pi\times 300\,\mathrm{Hz}$ from the other two, which is a difference small enough that we can ignore it. (We omit any other possible sources of energy shifts, such as level-dependent tensor shifts caused by optical traps.) The relevant molecular transition dipole moments at electric field $E_\mathrm{dc}^\mathrm{res}$ are found from numerical calculations, and they have the same value for transitions to all three states ${ |u_q^{(k)}\rangle }$:
\begin{equation}
 \mu_\updownarrow \equiv \langle u_q^{(k)}| \hat{d}_q |\downarrow^{(k)}\rangle \approx 0.561 d = 1.72\,\mathrm{debye}.
\end{equation}

\begin{table}[t]
\begin{tabular}{l@{\hskip 0.25in}l}
\hline
\hline
$E^\mathrm{res}_\mathrm{dc}$ & $3042.5\,\mathrm{mV/cm}\ddagger$ \Tstrut \Bstrut\\
\hline
$B_\mathrm{rot}$ & $2\pi\times10267.539\,\mathrm{MHz}$~\cite{1994-Anderson-APJ} \Tstrut\\
$d$ & $3.07\,\mathrm{debye}$~\cite{1984-Childs-JChemPhys} \\
$c_S$ & $2\pi\times20528.349\,\mathrm{MHz}$\\
$\mu_\updownarrow$ & $1.72\,\mathrm{debye}$ \Bstrut\\
\hline
$c_0$ & $2\pi\times20528.328\,\mathrm{MHz}\ddagger$ \Tstrut\\
$c_\Delta$ & $-2\pi\times21\,\mathrm{kHz}\ddagger$ \\
$\mu_\Updownarrow$ & $2539.79\,\mathrm{debye}\ddagger$ \Bstrut\\
\hline
$\mu_\updownarrow \mu_\Updownarrow/(4 \pi \epsilon_0 |1.5\,\mathrm{\mu m}|^3)$ & $2\pi\times195\,\mathrm{kHz}$ \Tstrut \Bstrut\\
\hline
Lifetime of ${ |\Uparrow\rangle }$ & $1.1 \times 10^{-4}\,\mathrm{s}\ddagger$ \Tstrut\\
Lifetime of ${ |\Downarrow\rangle }$ & $2.5 \times 10^{-4}\,\mathrm{s}\ddagger$ \Bstrut\\
\hline
\hline
\end{tabular}
\caption{Relevant physical parameters of the $^{85}$Rb--$^{40}$Ca$^{19}$F hybrid system in an electric field $E_\mathrm{dc} = E^\mathrm{res}_\mathrm{dc}$, tuned to equalize the transition frequencies $c_0,c_S$. $\ddagger$ marks atomic state properties obtained by a calculation with the Python \textsc{ARC} package. Radiative atomic state lifetimes are calculated for the $0\,\mathrm{K}$ temperature limit, under the assumption of no electronic or magnetic fields.}
\label{tab:CaFRbparameters}
\end{table}

The atomic pseudospin states are chosen as
\begin{align}
|\Downarrow\rangle &= \left|\overline{n=49,l=1,j=\frac{3}{2},m_j=+\frac{1}{2}}\right\rangle \\
|\Uparrow\rangle &= \left|\overline{n=48,l=2,j=\frac{5}{2},m_j=+\frac{3}{2}}\right\rangle.
\end{align}
The corresponding transition dipole moment is found numerically as $\mu_\Updownarrow \equiv {\langle \Uparrow | \hat{D}_{+1} |\Downarrow\rangle} = 2539.79\,\mathrm{debye}$.

Using Eq.~\eqref{eq:linear_combination_interaction_coefficient}, the expression for the resulting $C_k$ is
\begin{align}
C_k &= \sqrt{ \sum_{q=-1}^{+1} \left|\langle \Downarrow ; u_q^{(k)} | V_\mathrm{atom-mol}^{(k)} | \Uparrow ; \downarrow^{(k)} \rangle\right|^2} \nonumber \\
&= \sqrt{ \sum_{q=-1}^{+1} \left|\langle \Downarrow |\hat{D}_{-1} |\Uparrow\rangle \langle u_q^{(k)} | \hat{d}^{(k)}_q |\downarrow^{(k)}\rangle v^{(k)}_{-1;q}\right|^2} \nonumber \\
&= \sqrt{ \sum_{q=-1}^{+1} \left|- \mu_\Updownarrow \mu_\updownarrow v^{(k)}_{-1;q}\right|^2 } \nonumber \\
&= \frac{|\mu_\Updownarrow| |\mu_\updownarrow| }{4\pi\epsilon_0 |R_k|^3} \sqrt{\frac{5-3\cos^2\theta_k}{2}},
\end{align}
which at $R_k = 1.5\,\mathrm{\mu m}$ gives
\begin{equation}
 C_k = 2\pi\times195\,\mathrm{kHz} \sqrt{\frac{5-3\cos^2\theta_k}{2}}.
\end{equation}

Similarly to the example described in Section~\ref{sec:Model-ExampleChoice}, the electric field strength $E^\mathrm{res}_\mathrm{dc}$ has been found numerically (defined with accuracy to $0.1\,\mathrm{mV/cm}$) to minimize the resulting transition energy mismatch $c_\Delta = -2\pi\times 21\,\mathrm{kHz}$. Around the value $E^\mathrm{res}_\mathrm{dc}$, changing $E_\mathrm{dc}$ by $0.1\,\mathrm{mV/cm}$ changes $c_\Delta$ by approximately $2\pi\times65\,\mathrm{kHz}$. Therefore keeping $c_\Delta$ smaller than the interaction strength requires controlling the electric field strength down to the $\sim 0.1\,\mathrm{mV/cm}$ level. Such fine control is difficult, but plausible, since it is possible to measure (and subsequently compensate) stray electric fields with accuracy down to $\sim 0.01\,\mathrm{mV/cm}$~\cite{1999-Osterwalder-PRL,2014-Huber-NatComm}. The parameters of this example system are given in Table~\ref{tab:CaFRbparameters}.

\bibliography{_Biblio}

\end{document}